\documentclass{ar-1col}

\usepackage{natbib}

\def\aj{{AJ}}                   
\def\araa{{ARA\&A}}             
\def\apj{{ApJ}}                 
\def\apjl{{ApJ}}                
\def\apjs{{ApJS}}               
\def\aap{{A\&A}}                
\def\mnras{{MNRAS}}             
\def\pasa{{PASA}}               
\def\pasp{{PASP}}               
\def\nat{{Nature}}              
\def\ssr{{Space~Sci.~Rev.}}     

\jname{Xxxx. Xxx. Xxx. Xxx.}
\jvol{AA}
\jyear{YYYY}
\doi{10.1146/((please add article doi))}

\begin{document}

\markboth{Gal-Yam et al.}{Superluminous Supernovae}

\title{The Most Luminous Supernovae}

\author{Avishay Gal-Yam$^1$ 
\affil{$^1$Department of Particle Physics and Astrophysics, Weizmann Institute of Science, Rehovot, Israel, 76100; email: avishay.gal-yam@weizmann.ac.il}
}

\begin{abstract}
Over a decade ago, a group of supernova explosions with peak luminosities far exceeding (often by $>100$) those of normal events, has been identified. These superluminous supernovae (SLSNe) have been a focus of intensive study. I review the accumulated observations and discuss the implications for the physics of these extreme explosions. $\bullet$ SLSNe can be classified into hydrogen poor (SLSNe-I) and hydrogen rich (SLSNe-II) events. $\bullet$ Combining photometric and spectroscopic analysis of samples of nearby SLSNe-I and lower-luminosity events, a threshold of M$_{g}<-19.8$\,mag at peak appears to separate SLSNe-I from the normal population. $\bullet$ SLSN-I light curves can be quite complex, presenting both early bumps and late post-peak undulations. $\bullet$ SLSNe-I spectroscopically evolve from an early hot photospheric phase with a blue continuum and weak absorption lines, through a cool phtospheric phase resembling spectra of SNe Ic, and into the late nebular phase. $\bullet$ SLSNe-II are not nearly as well studied, lacking information based on large sample studies. Proposed models for the SLSN power source are challenged to explain all the observations. SLSNe arise from massive progenitors, with some events associated with very massive stars (M$>40$\,M$_{\odot}$). Host galaxies of SLSNe in the nearby universe tend to have low mass and sub-solar metallicity. SLSNe are rare, with rates $<100$ times lower than ordinary SNe. SLSN cosmology and their use as beacons to study the high-redshift universe offer exciting future prospects.    

\end{abstract}

\begin{keywords}
supernovae
\end{keywords}
\maketitle

\tableofcontents

\section{INTRODUCTION: A BRIEF HISTORICAL REVIEW}
The luminosity of supernovae (SNe) was the defining feature of these remarkable stellar events, and set these apart from less luminous and much more common stellar eruptions such as classical novae. Observed SN populations are invariably dominated by Type Ia events, the most luminous of all common SNe (e.g., \citealt{2013PASP..125..749G}). Analyzing the statistics of peak SN luminosities, \cite{2002AJ....123..745R} noted a population of rare, overluminous events that are significantly more luminous than SNe Ia (L$>1.2 \times 10^{43}$\,erg\,s$^{-1}$; M$<$-19.5). Following the discovery and detailed studies of a few nearby superluminous events (and in particular SN 2005ap, \citealt{2007ApJ...668L..99Q}; SN 2006gy, \citealt{2007ApJ...659L..13O}, \citealt{2007ApJ...666.1116S}; and SN 2007bi, \citealt{2009Natur.462..624G}) this population become a focus of intense study. In particular, \cite{2011Natur.474..487Q} used observations across a range of redshift in order to unify several fragmented reports and define a spectroscopic class of hydrogen-poor super-luminous events. \cite{2012Sci...337..927G} reviewed extant observation, provided a detailed historical review, set a fiducial threshold to consider an event as a superluminous supernova (SLSN), and introduced the subclasses of hydrogen poor (SLSN-I) and hydrogen-rich (SLSN-II) events. More recently, this class of objects was reviewed by \cite{2017hsn..book..431H} and by \cite{2017hsn..book..195G} in the context of supernova classification, while \cite{2018SSRv..214...59M} provide a broad theoretical perspective. The goal of this review is to update and expand on the work of \cite{2012Sci...337..927G}, using a substantially larger data set collected since: 98 SLSNe are currently listed on WISeREP \citep{2012PASP..124..668Y}, a sample $>5$ times larger than the 18 events available for analysis in 2012. 

\section{DEFINING THE CLASS OF SUPERLUMINOUS SUPERNOVAE}
\subsection{Luminosity Threshold}
\label{subsec_thresh}
SLSNe are by definition more luminous than normal events
(see \citealt{2017hsn..book..195G} for a summary of the properties
of normal SN classes), with individual events as luminous as L$_{\rm bolometric}$$>3\times10^{44}$\,erg\,s$^{-1}$ \citep{2014ApJ...797...24V}. \cite{2012Sci...337..927G} set an arbitrary
fiducial luminosity threshold of absolute magnitude L$>7\times10^{43}$\,erg\,s$^{-1}$ (M$<$-21\,mag) in any band to consider object as superluminous. The drawbacks of this choice (beyond its arbitrary nature) are obvious, and include ambiguity due to choice of observing bands, color effects and K-corrections, that are significant for SLSNe at high redshifts. Additional ambiguity involves the event timescale. In this review, we focus on events that maintain their peak luminosity over a minimal time scale of $>48$\,hours. This excludes shorter, luminous flares that can reach $>10^{44}$\,erg\,s$^{-1}$, such as those associated with supernova explosion shocks (e.g., \citealt{2017NatPh..13..510Y}) and Gamma-Ray Burst afterglow emission that is associated with some SNe Ic-BL. We also exclude events with a peak luminosity that is within the range of standard events, but that are exceptionally long-lived such that they can reach a high integrated luminosity over several years (e.g., \citealt{2010MNRAS.404..305M,2014ApJ...785...37B,2017Natur.551..210A}). 

\begin{marginnote}[]
Transient sky surveys providing samples of SLSNe: 
\entry{PTF}{The Palomar Transient Factory; \cite{2009PASP..121.1395L}}
\entry{PS1-MDS}{The PanSTARRS1
Medium Deep Survey (PS1 MDS); \cite{2016arXiv161205560C}}
\entry{TSS}{The Texas Supernova Survey (TSS); \cite{2013MNRAS.431..912Q}}
\entry{SNLS}{The Supernova Legacy Survey (SNLS); \cite{2017MNRAS.464.3568P}}
\end{marginnote}

\cite{2018ApJ...860..100D} inspected a large sample of Type I (hydrogen poor) SLSNe, as well as normal and broad-line (BL) SNe Ic from PTF in order to test whether a natural luminosity threshold emerges from the data (e.g., a double peaked luminosity distribution of normal and superluminous events). However, after correcting for the much larger volume probed by the PTF survey for SLSNe, the luminosity distributions of SLSNe-I, SNe Ic and SNe Ic-BL seem to join smoothly into a single distribution. Current observations therefore do not support a natural luminosity threshold separating SLSNe-I from lower-luminosity SNe Ic.

\begin{figure}[h]
\includegraphics[width=3in]{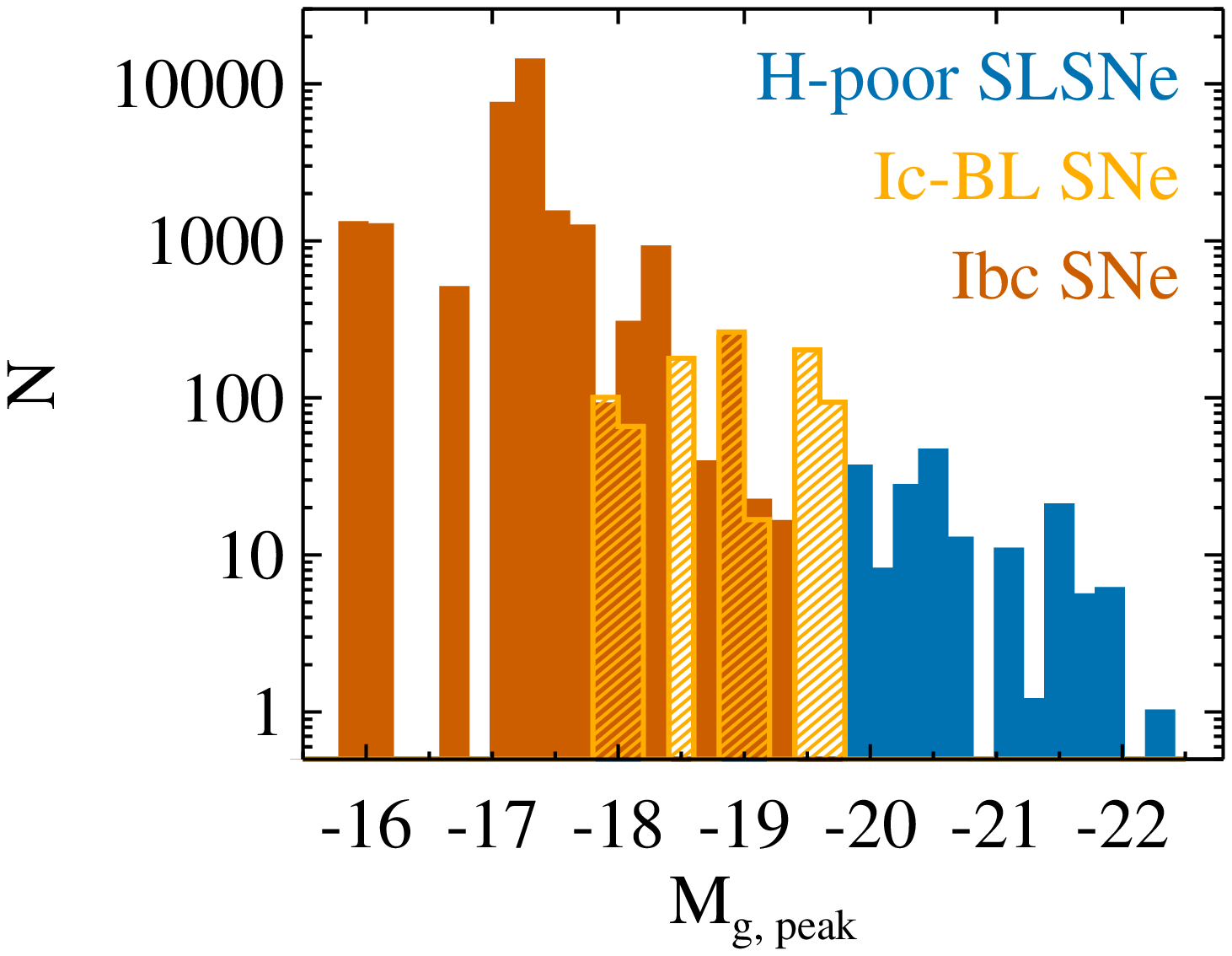}
\caption{A histogram of the luminosity of SLSNe-I, SNe Ic and SNe-Ic-BL from PTF with a volumetric correction accounting for the detectability of more luminous events out to larger distances, adapted from \cite{2018ApJ...860..100D} with perission from the AAS. The more luminous events (blue) also share spectral similarity according to \cite{2018ApJ...855....2Q} suggestion a threshold of M$_{g}=-19.8$\,mag may separate SLSNe-I from lower-luminosity events.}
\label{Fig_decia_hist}
\end{figure}

\cite{2018A&A...609A..83I} conduct a more sophisticated analysis of the photometric properties of SLSNe-I and propose a set of 4 photometric parameters (the peak magnitude at restframe 400\,nm, the magnitude decline in this band within 30 restframe days from peak, and the 400\,nm - 520\,nm color at peak and at 30\,d after), that can be used to statistically select SLSNe-I without a luminosity threshold. While this method could be used to select SLSNe-I from large samples of events, it has several limitations: It does not apply to all SLSNe-I (for example, those with double-peaked light curves); it is only $90\%$ complete even within the restricted event sub-sample presented by \cite{2018A&A...609A..83I}, it requires multiband photometric coverage during and after peak, and its selection purity (number of false SLSN identifications) has not been robustly estimated. It thus cannot be used as a definition of the SLSN-I class at this time.

Similar analysis is yet to be carried out for hydrogen-rich SLSNe-II. While individual events exist that span the range ($>5$\,mag) of luminosities from normal events to the most luminous SLSN-II, it is unclear whether the most luminous events are drawn from a luminosity distribution that smoothly extends to normal events.

\subsection{Spectroscopic Definition}
Lacking a clear luminosity threshold, one may hope to find a spectroscopic definition for SLSNe, i.e., a set of spectroscopic features that are common to SLSNe and differ from those of normal, lower-luminosity events. \cite{2018ApJ...855....2Q} were able to achieve this for the hydrogen poor (SLSN-I) variety. Applying an estimate for spectral similarity based on automated $\chi^{2}$ minimization ranking against spectral template libraries, these authors show that the most luminous events are also spectroscopically similar, and a threshold exists (M$_{\rm g}=-19.8$\,mag) that separates the more luminous events (blue in Fig.\ref{Fig_decia_hist}) and more common events with normal luminosity. As this threshold separates presumably two distributions of different object clasees, one does not expect it to be sharp: rare luminous ``normal'' SNe may have brighter peak magnitudes, as there could be some spectroscopic SLSNe-I that fall below this threshold. Yet, at this time this value seems to offer a useful threshold. For example, 
the implied lower luminosity threshold for SLSNe-I  includes all of the events spectroscopically identified by \cite{2018ApJ...852...81L} as SLSNe-I in their luminosity-unbiased analysis of events from the PS1-MDS survey.  

At this time, no similar investigations have been conducted for SLSNe-II.

\subsection{Relation to Normal Events}

\subsubsection{Hydrogen-poor SLSNe-I}
Even though the volume-corrected luminosity distributions of SLSNe-I, SNe Ic and SNe Ic-BL join smoothly (\citealt{2018ApJ...860..100D}; Fig.\ref{Fig_decia_hist}), several differences exist between these populations suggesting that SLSNe are not just the upper luminosity tail of SNe Ic, as supported also by spectroscopic analysis \citep{2018ApJ...855....2Q}. \cite{2018ApJ...860..100D} point out that the rise time of SLSNe-I is significantly longer compared to that of normal events with similar decay times, i.e., that the ratio of rise to decay time constitutes a photometric criterion to statistically separate these populations. In addition, as discussed below, early bumps prior to the first peak appear to be  common in SLSNe-I \citep{2016MNRAS.457L..79N} and are very rare in less luminous SNe Ic. Finally, \cite{2016ApJ...830...13P} and \cite{2018MNRAS.473.1258S} show that the fraction of SLSNe-I from all SNe is strongly suppressed in typically high-mass galaxies with metallicity above Z=0.4-0.5Z$_{\odot}$, unlike normal SNe Ic (but perhaps similar to SNe Ic-BL, e.g. \citealt{2010ApJ...721..777A,2012ApJ...759..107K}).     

\subsubsection{Hydrogen-rich SLSNe-II}
Due to very few studies, it is hard to determine whether and how SLSNe-II are related to lower-luminosity SNe II. Host galaxy properties do seem to suggest a distinction between SLSNe-II and lower luminosity SNe ($\S$~\ref{sec_hosts}). 

\begin{textbox}[h]\section{A photometric threshold for hydregen-poor superluminous supernovae (SLSN-I)}
The combination of photometric and spectroscopic analysis of the PTF sample of SLSN-I (\citealt{2018ApJ...860..100D,2018ApJ...855....2Q}) suggests an absolute magnitude threshold of {\bf M$_{g}=-19.8$\,mag}, above which all H- and He-poor events in this sample belong to a single spectroscopic similarity class that we identify as SLSN-I.
\end{textbox}
~

\section{CLASSIFICATION OF SUPERLUMINOUS SUPERNOVAE}

The classification scheme of SLSNe follows that of lower-luminosity events \citep{2017hsn..book..195G} and relies mainly on near-peak flux visible-light spectroscopy.

\subsection{SLSN-I}
\label{sec_class_I}

\begin{figure}[h]
\includegraphics[width=5in]{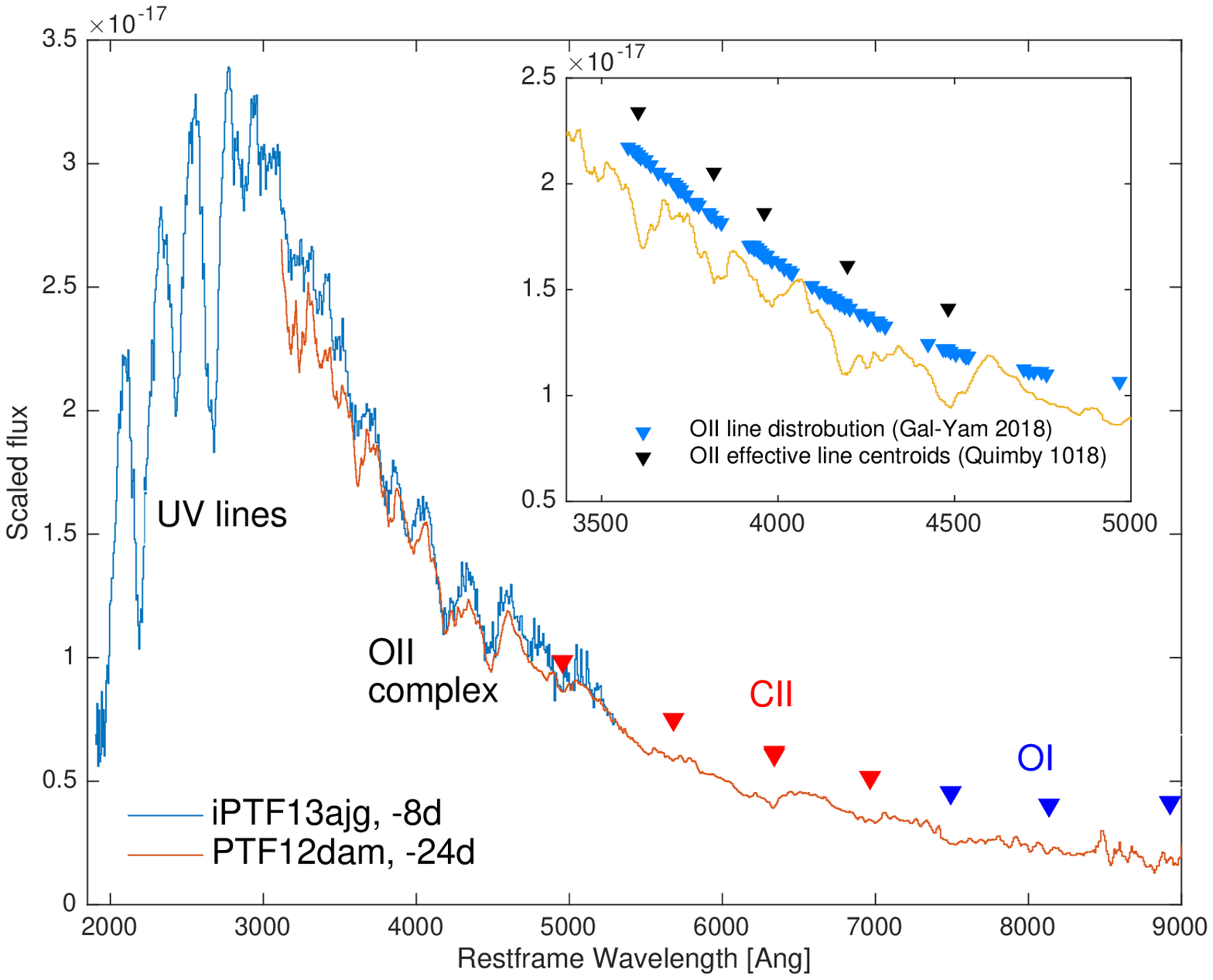}
\caption{Pre-peak spectra of SLSNe-I during the hot photospheric phase ($\S$~\ref{subsec_phot_hot}). Combining visible-light observations of typical high-redshift (iPTF13ajg, z=0.74; \citealt{2014ApJ...797...24V}) and low-redshift (PTF12dam, z=0.1075; \citealt{2018ApJ...855....2Q}) events traces the near UV and optical spectrum covering the SLSN emission peak. 
Weak lines of OI and CII are detected redward of 5000\AA. A dense forest of OII absorption features dominates the spectrum between 3500\AA\ and 5000\AA, with the strongest features appearing around 4200\AA\ and 4500\AA, and several strong absorption features are seen in the UV below 3200\AA. Inset: an illustration of the formation of the OII complex by a dense distribution of OII lines. Gaps in the distribution of OII transitions (blue) appear as emission ``peaks''. The effective central wavelengths of the five main OII absorption features calculated by \citealt{2018ApJ...855....2Q} (4650.71\AA, 4357.97\AA, 4115.17\AA, 3959.83\AA, and 3737.59\AA) are plotted in black and match the data very well. Lines of all elements have been shifted by an expansion velocity of 11000\,km\,s$^{-1}$.
}
\label{Fig_class_slsn_I}
\end{figure}

\begin{figure}[h]
\includegraphics[width=5in]{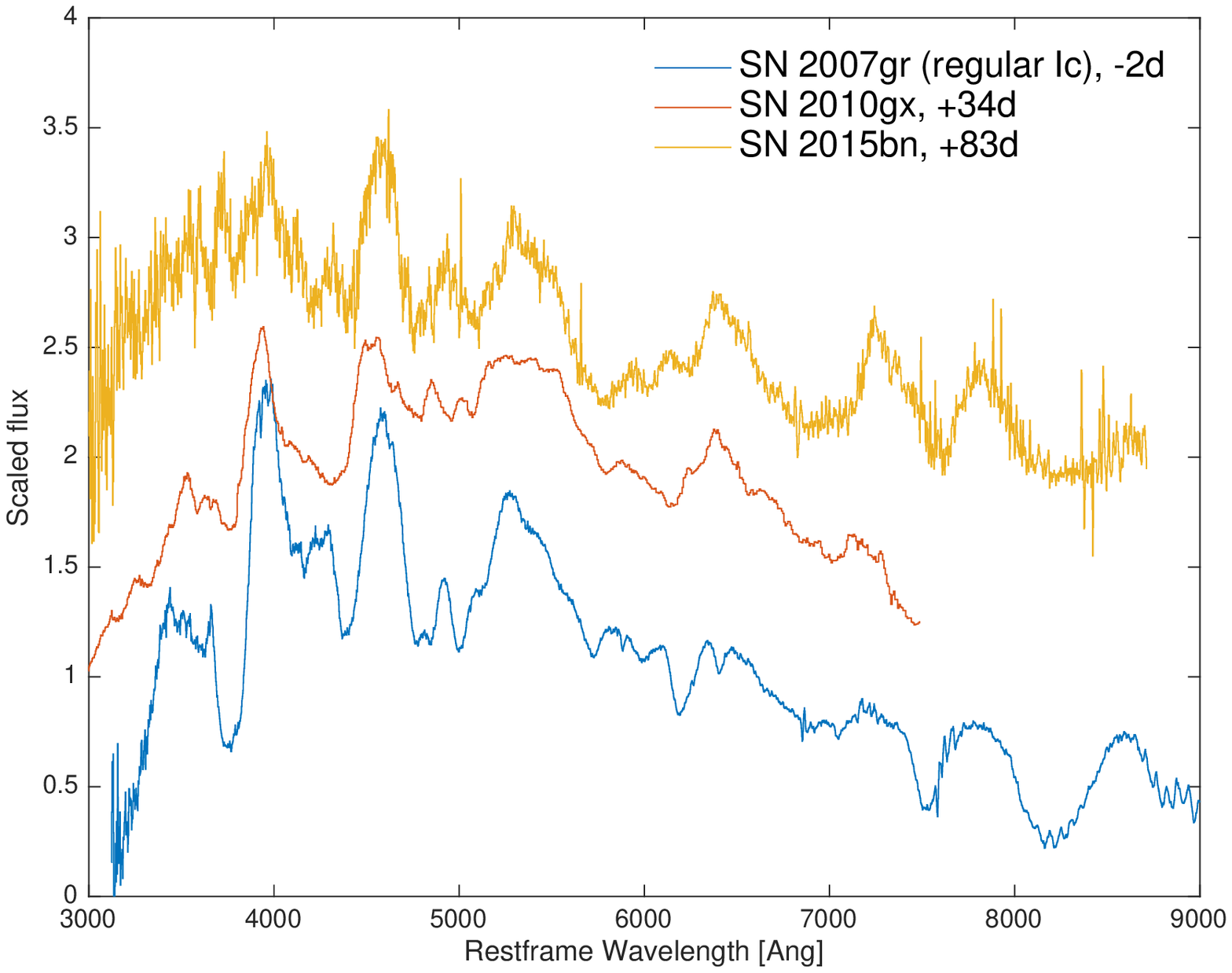}
\caption{Post-peak spectra of SLSNe-I during the cool photospheric phase ($\S$~\ref{subsec_phot_cool}). Several weeks (e.g., SN 2010gx; \citealt{2010ApJ...724L..16P}, \citealt{2018ApJ...855....2Q}) to months (e.g., SN 2015bn; \citealt{2016ApJ...826...39N}) after peak, SLSN-I evolve to resemble normal SNe Ic around peak (SN 2007gr shown for comparison; \citealt{2008ApJ...673L.155V}). 
}
\label{Fig_class_slsn_I_late}
\end{figure}

Fig.\ref{Fig_class_slsn_I} shows typical pre-peak spectra of this class of hydrogen-poor SLSNe. The strongest lines in visible-light spectra of SLSNe-I arise from the cumulative absorption of a large number of dense OII transitions in the blue (3000\AA\ $<\lambda<$ 5000\AA; see, e.g., \citealt{2018ApJ...855....2Q}). In the UV there are several strong absorption features, while the red part of the spectrum shows typically much weaker features of OI and CII (see, e.g., \citealt{2018arXiv180608224G}). At later phases, weeks to months after peak, the spectrum evolves and becomes similar to that of more normal SNe Ic around peak (Fig.\ref{Fig_class_slsn_I_late}). Below we refer to these two phases as the hot and cool photospheric phases (see $\S$~\ref{subsec_phot_hot} and $\S$~\ref{subsec_phot_cool}, respectively). 

Spectroscopic classification of SLSNe-I is thus usually done at or before peak based on either the identification of the OII complex in low-redshift events, or the UV lines that are redshifted into the visible for higher-redshift SLSNe (Fig.\ref{Fig_class_slsn_I}). Events lacking peak spectra can alternatively be classified by a combination of high peak luminosity and late-time similarity to SNe Ic (Fig.~\ref{Fig_class_slsn_I_late}).  

\begin{marginnote}[8cm]
Spectra shown in this and all other figures are publicly available from WISeREP \citep{2012PASP..124..668Y}
\end{marginnote}

\subsubsection{Proposed subdivisions}
\cite{2012Sci...337..927G} proposed a possible division among hydrogen-poor SLSNe into rapidly declining SLSNe-I and slowly-declining events (with a decay rate consistent with radioactive decay of $^{56}$Co) designated SLSN-R. This division was not broadly adopted. Recent sample studies (e.g., \citealt{2018ApJ...860..100D}, \citealt{2018ApJ...852...81L} and \citealt{2017ApJ...850...55N}) fail to find a clear natural division into two distinct classes of rapidly- and slowly-declining hydrogen-poor events (though \citealt{2018ApJ...852...81L} may see a hint for bimodality). 

\cite{2018ApJ...854..175I} conduct a more sophisticated statistical analysis (including unsupervised clustering) of multiple photometric and spectroscopic parameters of SLSNe-I, 
and claim that a restricted set of events does cluster into two groups defined by rapid/slow evolution. This interesting work is limited by small numbers (only 3 events cluster into the slow category). \cite{2017MNRAS.468.4642I} inspect the light curve properties of slowly-fading SLSNe-I, and propose that these may show ubiquitous post-peak undulations. Spectroscopic analaysis conducted by \cite{2018ApJ...855....2Q} also suggests a division based on similarity to two prototype events (PTF12dam and SN2011ke) that are slowly- and rapidly-declining SLSNe-I. A clear test of whether distinct groups of SLSN-I exist would greatly benefit from the large numbers of events expected for forthcoming massive surveys. Until that time, the term SLSN-I is applied to the entire class.

\subsection{SLSN-II}

Hydrogen-rich SLSNe-II show clear spectroscopic signatures of hydrogen around peak. He features are often also seen, especially in early, hot events. 

\begin{figure}[h]
\includegraphics[width=5in]{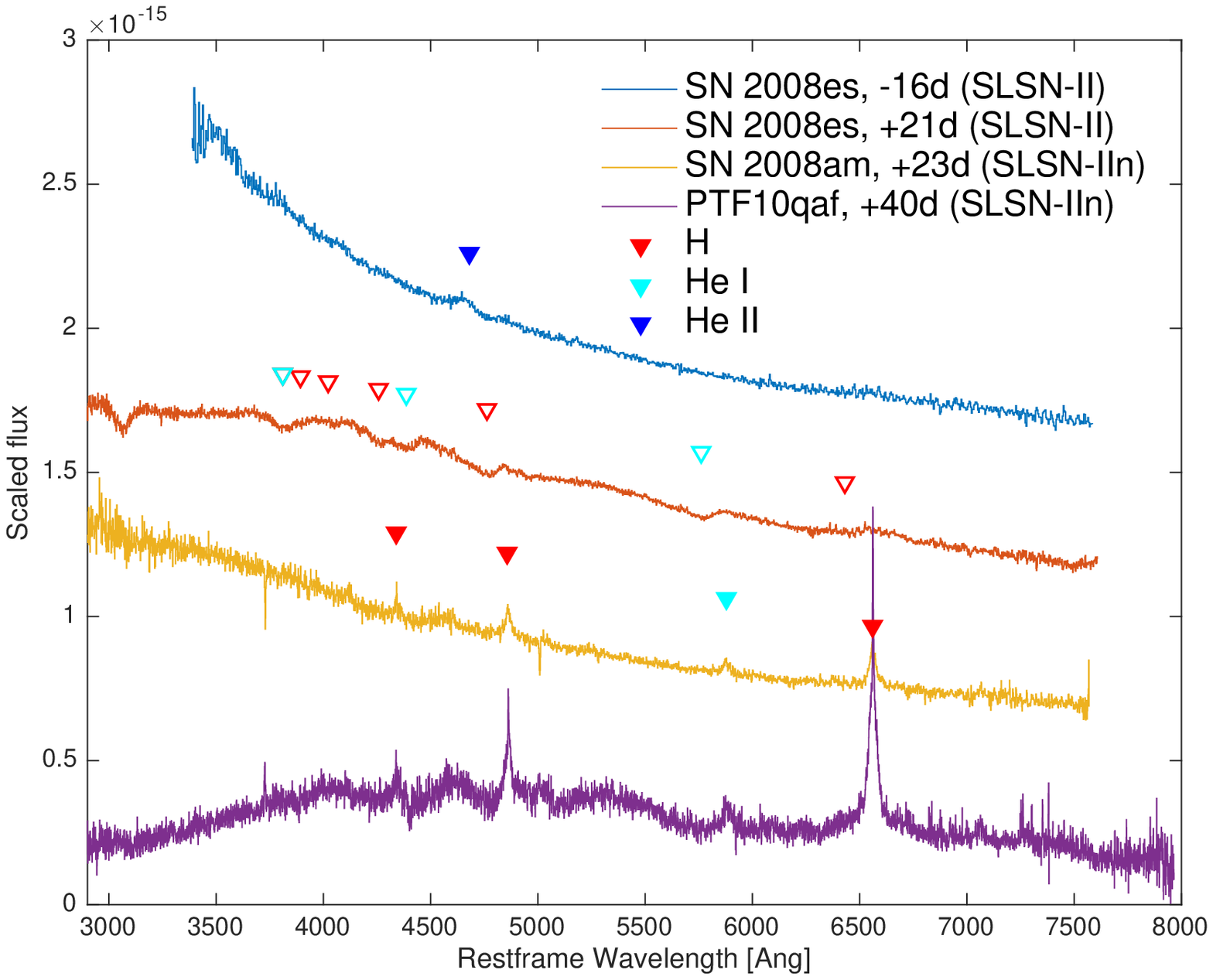}
\caption{Spectra of SLSNe-II. SLSNe-IIn show strong H Balmer emission lines over a blue continuum at early time (SN 2008am; \citealt{2011ApJ...729..143C}) evolving into a more complex spectrum after peak (PTF10qaf; \citealt{2012Sci...337..927G}). Rare SLSNe-II do not show narrow emission lines but rather blue continuum spectra (with broad He II in the case of SN 2008es shown; \citealt{2009ApJ...690.1313G}) that evolve into broad absorption features of H and He I after peak (\citealt{2009ApJ...690.1303M}; empty markers have been blueshifted by 6000 km\,s$^{-1}$ to match the observed absorption features).
}
\label{Fig_class_slsn_II}
\end{figure}

\subsubsection{The Narrow-line SLSN-IIn class}

The majority of events (similar in numbers to SLSNe-I; \citealt{2012Sci...337..927G}; \citealt{2016ApJ...830...13P}) show prominent emission lines of the hydrogen Balmer series, often accompanied by He I, with a narrow core over a broader base (Fig.~\ref{Fig_class_slsn_II}), similar to lower luminosity SNe IIn. Initially some objects show a blue continuum with Balmer emission lines only, while more complex spectra develop later (Fig.~\ref{Fig_class_slsn_II}).

\subsubsection{The SLSN-II class and its relation to SLSN-I}
\label{subsubsec_slsn_II_I}

The first hydrogen-rich SLSN that did not exhibit narrow Balmer lines was SN 2008es (\citealt{2009ApJ...690.1313G}; \citealt{2009ApJ...690.1303M}; Fig.~\ref{Fig_class_slsn_II}). The sample of similar events remains small and currently includes but a handful of additional events including CSS121015 \citep{2014MNRAS.441..289B}, SN 2013hx and PS15br \citep{2018MNRAS.475.1046I}.  

The group is characterized by spectra
with evident Balmer lines in absorption or with P-Cygni profiles, but lacking narrow IIn-like features, an important distinction as such narrow features are generally taken as evidence for strong CSM interaction. Spectra are typically featureless initially with lines developing with time and becoming strong post-peak. Late spectra, in particular those studied by \cite{2014MNRAS.441..289B} evolve to strongly resemble late SLSN-I spectra superposed with a weak Balmer series, suggesting a connection between these rare SLSNe-II and the more common SLSNe-I, perhaps driven by a small amount of residual hydrogen left in the outer stellar envelope when these SNe explode.

PTF10hgi is currently a unique case of a SLSN that was initially classified as a SLSN-I, but actually shows clear spectral
signatures of hydrogen and helium \citep{2018ApJ...855....2Q}, 
motivating a classification of SLSN-IIb in analogy to lower luminosity events. Interestingly, this event is also among the few that cannot be satisfactorily fit by a parameterized magnetar model by \cite{2018ApJ...860..100D}, and is an outlier in the analysis of \cite{2015MNRAS.452.3869N}, having very slow photospheric velocities.

\begin{textbox}[h]\section{Summary of SLSN classification}
Hydrogen-poor SLSNe-I lack strong features of hydrogen in peak-light spectra. Hydrogen-rich events include SLSNe-IIn that show strong narrow emission lines of hydrogen, SLSNe-II that show broad hydrogen absorption features, and a single SLSN-IIb that shows strong absorption features of both H and He.  
\end{textbox}
~

\section{OBSERVED PROPERTIES OF SLSNE}
\subsection{Photometry}
\begin{figure}[h]
\includegraphics[width=3.6in]{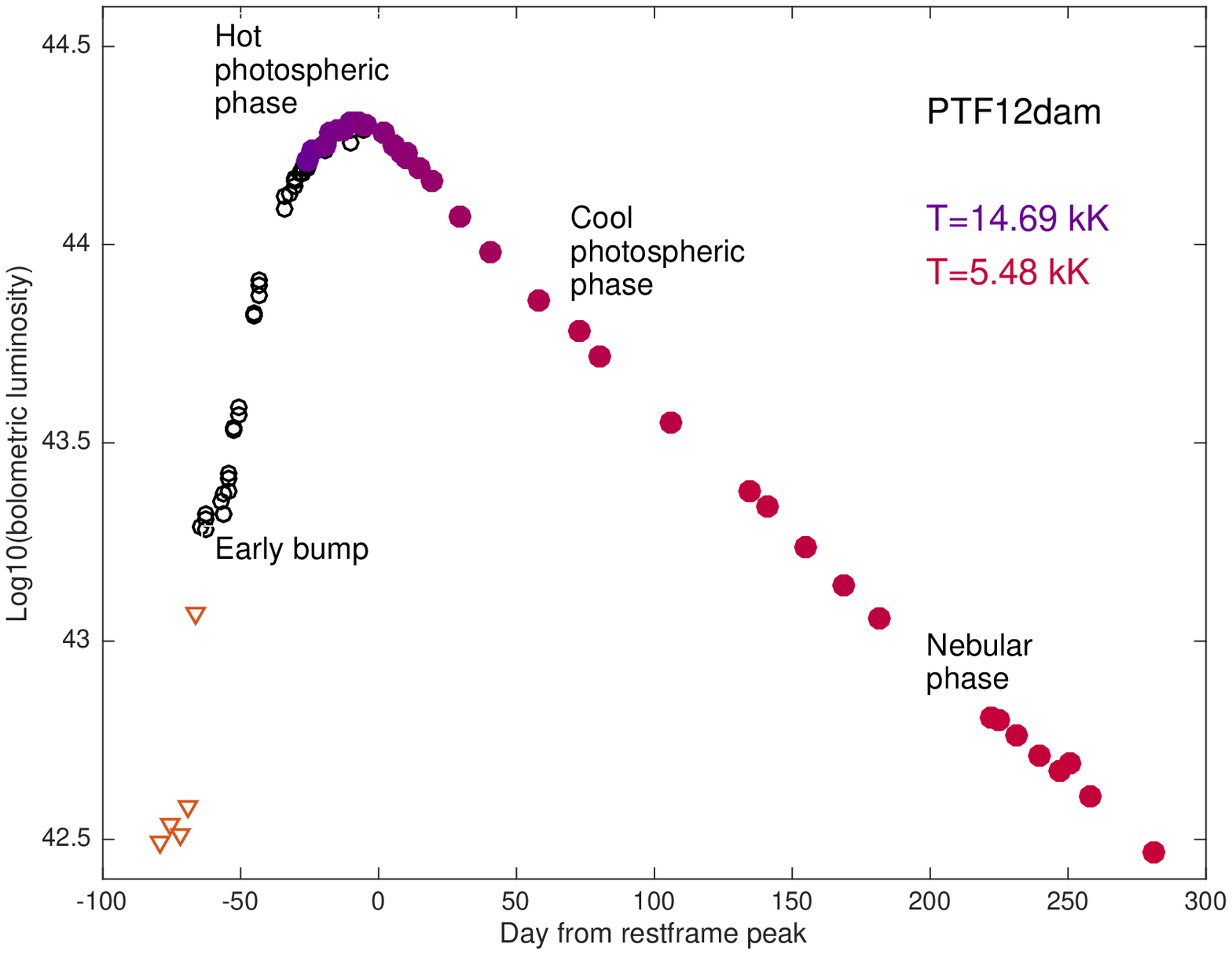}
\includegraphics[width=3.6in]{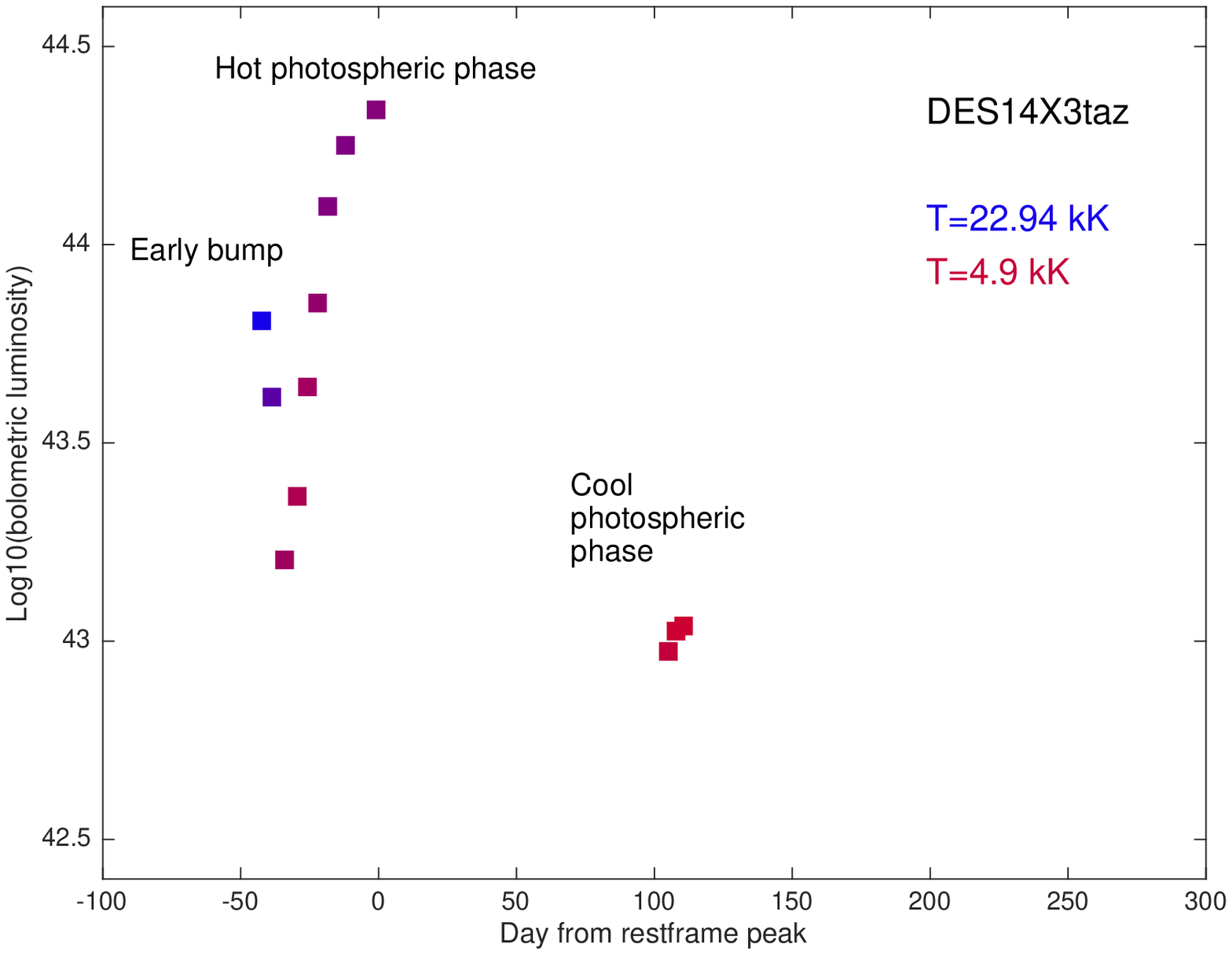}
\caption{Top: the bolometric light curve of PTF12dam, a well-observed slowly-declining SLSN-I from \cite{2017ApJ...835...58V}. The symbols are color-coded based on the temperature fits from \cite{2017ApJ...835...58V}; values prior to -25\,d require extrapolations and are less certain, and not coded. Main spectroscopic phases discussed below are marked for reference. PTF12dam transitioned from the hot to the cool photospheric phase about a month after peak, and went into the nebular phase around 200\,d post-peak. 
An early bump is detected shortly after explosion \citep{2017ApJ...835...58V}.
Late (post-peak) bumps, often seen in slowly-declining events \citep{2017MNRAS.468.4642I} are not obvious in the bolometric light curve of this event.
Bottom: A similar plot for the rapidly-declining SLSN-I DES14X3taz from \cite{2016ApJ...818L...8S}.
This event has a prominent early bump, during which the temperature measured is significantly hotter than at peak. The decline from peak to the cool photospheric phase is rapid but not well measured; for such rapid decliners, nebular phase observations are challenging.
}
\label{Fig_LC}
\end{figure}

As a class, SLSNe of both types I and II typically have longer timescales in both rise and fall from peak compared to most lower luminosity events. Fig~\ref{Fig_LC} shows the bolometric light curve of PTF12dam from \cite{2017ApJ...835...58V}. This slowly-declining SLSN-I was discovered shortly after explosions and extensively followed.

\subsubsection{SLSNe-I}
\paragraph{Typical light curves: peak}
As explained in $\S$~\ref{subsec_thresh}, when correcting for the larger volume within which more luminous events can be observed, SLSNe-I do not seem to separate from normal SNe Ib/c into a bimodal luminosity distribution, but hydrogen-poor SNe brighter than M$_{g}=-19.8$\,mag appear to belong to the spectroscopic class of SLSNe-I (\citealt{2018ApJ...860..100D,2018ApJ...855....2Q}). The range of peak magnitudes observed extends from this lower cutoff up to M$_g=-22.5$ or so for the brightest events (e.g., \citealt{2014ApJ...797...24V}). The mean peak luminosity of observed SLSN-I samples somewhat depends on the survey properties. Using a relatively small sample based on TSS, \cite{2013MNRAS.431..912Q} find a a typical peak magnitude of M$_{\rm unfiltered}=-21.7\pm0.4$\,mag. Several literature compilation studies find similar values, including \cite{2016MNRAS.457L..79N} that derived M$_{g}=-21.62\pm0.59$\,mag (assuming M$_{griz}=$M$_g-0.9$\,mag) and \cite{2018ApJ...854..175I} that measure M$_{400nm}=-21.7\pm0.5$\,mag. Applying a volumetric correction, \cite{2018ApJ...860..100D} find a mean g-band absolute peak magnitude of M$_{g}=-21.14\pm0.75$\,mag for the PTF sample, while \cite{2018ApJ...852...81L} measure M$_{400nm}=-21.1\pm0.7$\,mag from PS1. 
Transforming either observed or calculated restframe magnitude distributions into luminosities depends on assumptions regarding SLSN-I K-corrections, but typical luminosity values are $2.5\times10^{43}-5\times10^{44}$\,erg\,s$^{-1}$ (\citealt{2018ApJ...860..100D}, \citealt{2018ApJ...852...81L}). 

Of note is ASASSN-15lh (SN 2015L; \citealt{2016Sci...351..257D}) a transient with a peak magnitude that stands out above even the most luminous SLSNe-I (M$\sim$-23.5\,mag). However, the association of this transient with the class of SLSNe-I has been challenged \citep{2016NatAs...1E...2L,2018arXiv180608224G} and remains uncertain. 

\paragraph{Typical light curves: rise times}
Using the time difference between first detection and peak as a lower limit on the rise time, the SLSN-I samples of \cite{2018ApJ...852...81L} and \cite{2018ApJ...860..100D} include events with a bolometric rise time ranging from $\sim20$ to $>100$ restframe days, with PTF10nmn and PS1-14bj being examples of events with extremely long rise times from each sample, respectively. Even the fastest-rising events have longer bolometric rise times than those of normal hydrogen-poor type I SNe ($<20$\,d; e.g., \citealt{2016MNRAS.458.2973P}). Various quantitative measures of the rise time are provides in the literature. \cite{2018ApJ...860..100D} tabulate the time to rise by 1\,mag or from half flux to peak in restframe g band, while \cite{2015MNRAS.452.3869N} and \cite{2018ApJ...852...81L} tabulate the bolometric time to rise from 1/e below maximum to peak. \cite{2017ApJ...850...55N} provide bolometric rise times estimated from light curve models that are consistent with the values cited above.   

\paragraph{Typical light curves: decay times and slopes}
SLSN-I decay times are typically slower than those of normal SNe I (e.g., \citealt{2015MNRAS.452.3869N}, \citealt{2018ApJ...860..100D}). Since SLSNe are rarely (if ever) followed till their physical disappearance (i.e., when the SN is less luminous than its progenitor star), absolute decay times are not well defined. Various works tabulate decay slopes, ranging in restframe g band from $<0.01$ to about 0.08\,mag\,d$^{-1}$ (e.g., \citealt{2018ApJ...860..100D}), or times to decay by 1\,mag or to 1/2 of peak flux \citep{2018ApJ...860..100D} or by 1/e (\citealt{2015MNRAS.452.3869N}, \citealt{2018ApJ...852...81L}). Alternatively, measures of the light decay in a certain band in magnitudes within a given period in days (e.g., $\Delta$M$_{20}$, $\Delta$M$_{30}$) is measured, for example by \cite{2014ApJ...796...87I} and \cite{2018A&A...609A..83I}.   

\paragraph{Early bumps}
\label{early_bumps}
\cite{2016NatAs...1E...2L} identified an early $\sim10$\,d long bump in the rising light curve of the SLSN-I SN 2006oz. Additional examples were later identified including the well observed LSQ14bdq \citep{2015ApJ...807L..18N} and DES14X3taz \citep{2016ApJ...818L...8S}. \cite{2017ApJ...835...58V} presented a weak bump that was evident in early observations of PTF12dam, as well as an extreme early bump in the light curve of iPTF13dcc. Early bumps for which color information exists indicate the emission is initially very hot ($\sim25000$\,K) and cools rapidly \citep[see Fig.~\ref{Fig_LC}]{2016ApJ...818L...8S,2016MNRAS.457L..79N}. Bump durations range from about 10\,d to several weeks, and bump luminosities extend from about -19\,mag to above -21\,mag for the extreme iPTF13dcc, making these bumps brighter than almost all normal supernovae and in some cases rivaling in flux and duration the main peak of SLSNe. \cite{2016MNRAS.457L..79N} undertook a systematic search for such early bumps in the light curves of a literature sample of SLSNe-I. They find evidence for plausible early bumps in 8 of the 14 events with relevant early data, and can not rule out the existence of such a bump in any of the events, suggesting that these bumps may be common or even ubiquitous in SLSNe-I. However, the recent, nearby SN 2018bsz seems to lack a distinct bump, showing instead a long, slowly-rising early plateau \citep{2018arXiv180610609A}; and the sample of SLSNe from the DES survey also shows that significant early bumps are not ubiquitous (C. Inserra, private communication).  

\paragraph{Late bumps}
\label{late_bumps}
SN 2007bi, one of the first published SLSNe-I \citep{2009Natur.462..624G}, showed apparent minor undulations in its declining post-peak light curve. \cite{2016ApJ...826...39N} studied the well observed SN 2015bn, and documented several undulation occurring up to $\sim100$ days post peak, with typical durations of $\sim20$\,d and amplitudes of $\sim20\%$ compared to a smooth fit to the bolometric light curve. The bumps appeared to result from excess blue light. \cite{2017MNRAS.468.4642I} inspected several slowly-declining SLSNe-I, and found evidence for such small late-time bumps in all the objects studied. Light curves of more rapidly-declining events lack similar evidence for multiple bumps, though a single rapidly-declining event may show a late-time bump \citep{2016ApJ...826...39N,2017MNRAS.468.4642I} and in any case the rapid fading of these events makes detection of weak features post-peak more difficult. Whether such late bumps are correlated with a slowly-declining light curve thus remains an open question.

\subsubsection{SLSNe-II}
The information about SLSNe-II is scarce compared to studies of SLSNe-I. Preliminary results from a sample study of PTF events (Leloudas et al., in preparation, private communication) indicate that the sample, which is dominated by SLSNe-IIn, has a mean peak magnitude of M$_{r}=-21.1\pm0.5$. Rise times are comparable to, and decay times are longer than, those measured for SLSNe-I from the same survey. 

\paragraph{Precursors}
\label{slsn_II_precursors}
\cite{2014ApJ...789..104O} present the detection of a pre-explosion precursor eruption prior to the luminous Type IIn supernova 
PTF10weh (peak magnitude M$_{g}=-20.7$\,mag). It is not clear at this time whether there is a threshold separating SLSNe-IIn from lower-luminosity events, and if so, whether PTF10weh should be considered a SLSN-IIn. In any case, pre-explosion LBV-like massive eruptions could explain the evidence for massive CSM surrounding at least some SLSNe-IIn (see $\S~$\ref{subsubsec_csm} below).
\label{subsubsec_II_phot}

\subsection{Spectroscopy: SLSNe-I}
SLSNe-I evolve through three main phases in their spectral evolution.

\subsubsection{Hot photospheric phase}
\label{subsec_phot_hot}
Initially, SLSN spectra are dominated by a hot blue continuum, with black-body temperatures often reaching 20000\,K before maximum light (e.g., \citealt{2016ApJ...818L...8S,2018ApJ...853...57B}; Fig.~\ref{Fig_LC}). Absorption features are detected on top of the continuum, typically of OI and CII in the red part of the visible light spectrum (e.g., \citealt{2018arXiv180608224G}, \citealt{2018arXiv180610609A}), the OII absorption complex in the blue part of the visible spectrum (e.g., \citealt{2018ApJ...855....2Q}, \citealt{2018arXiv180608224G}) and several strong UV absorption features, see $\S$~\ref{sec_class_I}, $\S$~\ref{subsec_UV} and Fig.~\ref{Fig_class_slsn_I} above. 

The OII complex is quite unique to SLSNe-I, that are the only type of SNe where these features are detected persistently up to peak. As shown by 
\cite{2016MNRAS.458.3455M}, the formation of these lines requires population of highly excited levels of OII, which in turn implies significant departure from LTE and high temperatures; these conditions probably do not persist in normal hydrogen-poor SNe. OII has a very large number of lines in the blue part of the visible light spectrum, all with comparable strength. \cite{2018arXiv180608224G} show that gaps in the distribution of these lines appear as emission ``peaks'' in the spectrum (Fig~\ref{Fig_class_slsn_I}, inset). \cite{2018ApJ...855....2Q} synthesized the OII absorption spectrum based on line lists from NIST, and compared it to their data. They found some inconsistencies that might be due to a few specific poorly measured laboratory transition data, and after adjusting for these findings derive effective central wavelengths for the main absorption features that match the observations very well (Fig~\ref{Fig_class_slsn_I}, inset). 

The identification of the UV lines is less certain. 
As seen in Fig.~\ref{Fig_class_slsn_I}, there are four absorption features in the near UV, approximately around 2680\AA, 2450\AA, 2200\AA\ and 1950\AA. \cite{2018ApJ...855....2Q} review the proposed identifications of these lines by several studies \citep{2011Natur.474..487Q,2012MNRAS.426L..76D,2013ApJ...779...98H,2016MNRAS.458.3455M} that did not converge to a consensus. Adopting the identifications of \cite{2016MNRAS.458.3455M} whose models recover the spectral evolution of SLSNe-I across this and later phases, the reddest feature at 2680\AA\ is due to MgII and CII, The next two arise from a combination of TiIII and CII, along with SiIII (2450\AA) and CIII (2200\AA) while the bluest feature at 1950\AA\ is due to Fe III and Co III.

Typical expansion velocities derived from C and O lines before and around peak are v=$10000-15000$\,km\,s$^{-1}$ \citep{2018ApJ...855....2Q,2018arXiv180608224G}. The ejecta velocity dispersion $\Delta$v during this phase is remarkably narrow \citep{2018ApJ...855....2Q,2018arXiv180608224G} with lines from single transitions suggesting $\Delta$v=1500\,km\,s$^{-1}$ \citep{2018arXiv180608224G}, making SLSNe-I strong outliers to the correlation found by \cite{2016ApJ...832..108M} between expansion velocity (line blueshift) and velocity dispersion (line width) for SNe Ib/c \citep{2018arXiv180608224G}. 

Theoretical spectral synthesis studies of this phase have been conducted by \cite{2012MNRAS.426L..76D}, and more recently  \cite{2016MNRAS.458.3455M} developed spectral synthesis models that reproduce the evolution of SLSNe-I through this and later stages.   

\subsubsection{Cool photospheric phase}
\label{subsec_phot_cool}
As the photosphere expands and cools the spectrum of SLSNe-I evolves. In the visible, OII features weaken as the temperature falls below 12000\,K or so \citep{2018ApJ...853...57B,2016MNRAS.458.3455M}, and the spectrum rapidly changes to resemble spectra of more normal SNe Ic once the temperatures are below 10000\,K (Fig.\ref{Fig_class_slsn_I_late}; \citealt{2010ApJ...724L..16P}, \citealt{2011Natur.474..487Q}). \cite{2016MNRAS.458.3455M} predict that He I lines should briefly appear during the transition phase. This transition typically occurs a few days to weeks after peak, though some events (in particular some slowly evolving events like SN 2015bn; \citealt{2016ApJ...826...39N}) cool to the transition temperature well before peak magnitude. \cite{2017ApJ...845...85L} conducted a thorough investigation of the spectroscopic evolution of SLSNe-I in comparison with SNe Ic. They find that in terms of expansion velocities deduced from Fe II lines that become observable as the SNe cool, SLSNe-I are more similar to SNe Ic-BL than to normal SNe Ic.   

\subsubsection{Nebular emission}
As SNe continue to cool and expand, their ejecta eventually become transparent and they enter the nebular phase. This phase begins later in SLSNe ($>100$\,d post-peak) compared to normal SNe Ic, indicating high density is maintained in the ejecta during a longer period, due to more massive ejecta, ionization effects, or a combination of these (e.g., \citealt{2015MNRAS.452.3869N}).
Analysis of nebular spectra is a powerful probe of the physics of SNe, and it has been applied to SLSNe-I with suitable data over the years (e.g., \citealt{2009Natur.462..624G}, \citealt{2016ApJ...828L..18N}, \citealt{2017ApJ...835...13J})  

\begin{figure}[h]
\includegraphics[width=5in]{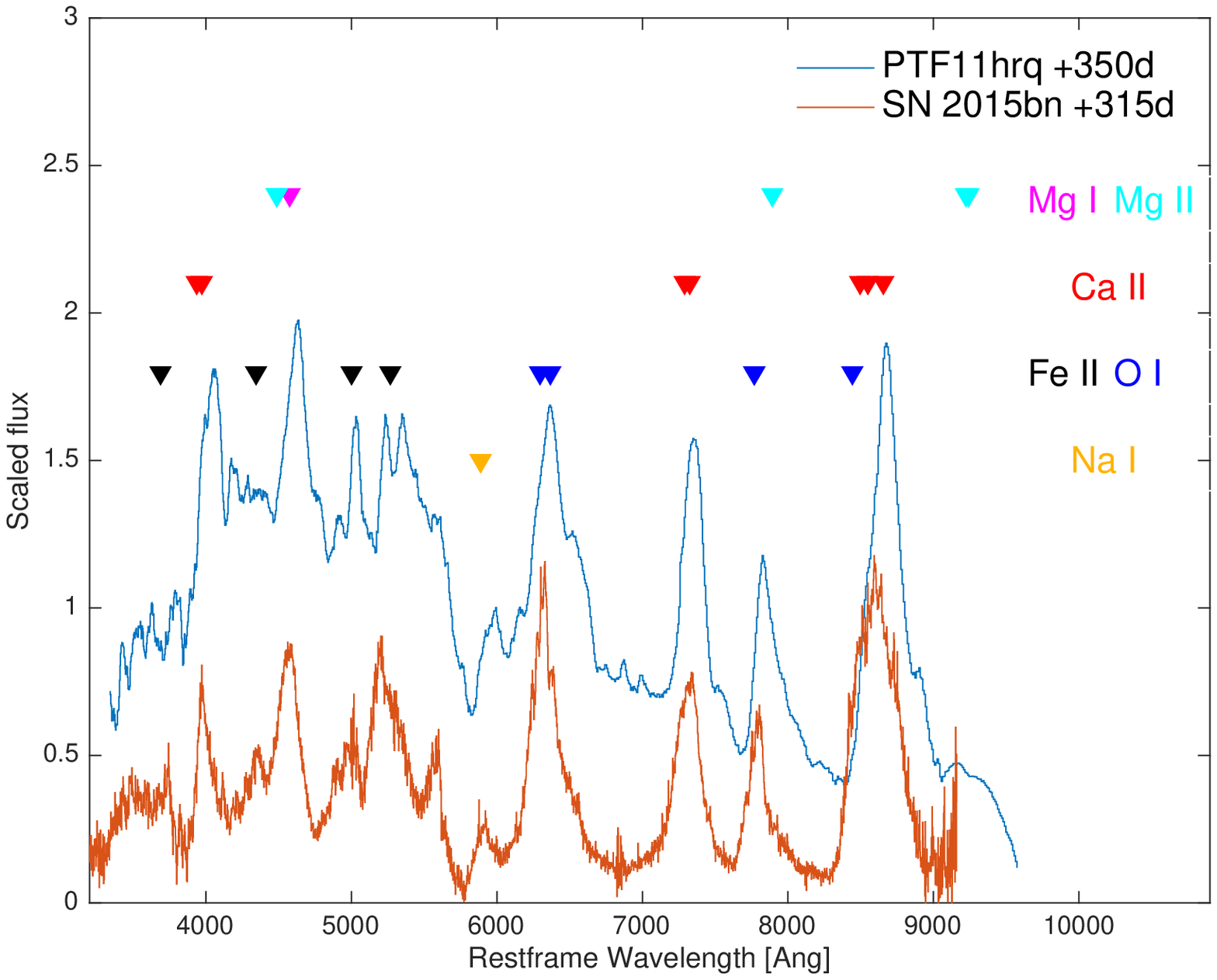}
\caption{Nebular spectra of SLSNe-I. High S/N spectra of 
PTF11hrq from \cite{2018ApJ...855....2Q} and SN 2015bn from \cite{2017ApJ...835...13J} are shown;
these are among the most nearby SLSNe-I with high quality late-time spectra. Line identifications are drawn from NIST; with
the addition of the semi-forbidden line of Mg I at 4571\AA.
}
\label{Fig_slsn_nebular}
\end{figure}

\cite{2018arXiv180800510N} combine new nebular phase observations with a large literature sample (in particular the large PTF data set released by \citealt{2018ApJ...855....2Q}) of SLSN-I spectra to conduct a detailed analysis of the nebular emission from these events, based on line identifications from \cite{2017MNRAS.468.4642I} and \cite{2017ApJ...835...13J}. In order to account for the intrinsically different timescales of SLSNe, they normalize the spectral phase according to the respective light curves. These authors define an early nebular phase, beginning two exponential decay timescales after peak (a mean restframe phase of $\sim180$\,d after peak; equivalent to $\times2.5$ the time to decline by 1\,mag measured by \citealt{2018ApJ...860..100D}), and a late nebular phase beginning after 4 exponential timescales ($\sim360$ restframe days on average) after peak. Early nebular spectra contain residual continuum emission and the strongest emission line is often Ca II 7300\AA, while late nebular spectra are typically dominated by the OI 6300\AA\ line (Fig.~\ref{Fig_slsn_nebular}). 

Strong emission lines in high S/N mean spectra calculated by \cite{2018arXiv180800510N} can be identified with ions of O, Mg, Ca, Fe and Na (Fig.~\ref{Fig_slsn_nebular}). The spectra are overall similar to those of SNe Ic, but show stronger emission from Fe in the blue (as noted early on by \citealt{2009Natur.462..624G}) suggesting a larger fraction of Fe-group elements in the ejecta. OI 7774\AA\ is stronger in SLSNe compared to SNe Ic, and in some events OII and OIII lines are seen \citep{2014ApJ...787..138L, 2017MNRAS.468.4642I}, indicating high ionization. Nebular line widths are remarkably high, with FWHM velocities of OI 6300\AA\ extending to above 15000\,km\,s$^{-1}$ at the early nebular phase, and dropping to $\sim8000$\,km\,s$^{-1}$ during the late phase.           

\subsubsection{Late-time emergence of hydrogen in SLSN-I}
\label{late_H}

\cite{2015ApJ...814..108Y} reported the first detection of an emerging emission of hydrogen in late-time spectra of the SLSN-I iPTF13ehe. \cite{2017ApJ...848....6Y} presented two additional cases of such SLSNe-I with late-time hydrogen emission, and discussed the properties of this small sample. The first spectra of iPTF13ehe were obtained around peak, and the object was already in the cool photospheric phase; it resembled slowly-evolving events like SN 2007bi and SN 2015bn in both its light curve and its early spectra. About 250\,d after peak, broad emission lines emerge, initially hydrogen H$\alpha$ and later (around 322\,d after peak) also H$\beta$ and He I 5876\AA\ become apparent. The two additional events presented by \cite{2017ApJ...848....6Y} have narrower light curves, and their earlier spectra (obtained around peak) do not resemble the standard hot photospheric spectra of SLSNe-I described in $\S$~\ref{subsec_phot_hot} \citep{2018arXiv180608224G}, suggesting perhaps that SLSNe-I that present these late-time interaction features may also have distinct early spectra. 
The observations require a distribution of hydrogen-rich material lying at distances of a few$\times10^{16}$\,cm. \cite{2015ApJ...814..108Y} estimate $\sim15\%$ of SLSNe-I may show late-time hydrogen Balmer emission.

\subsection{Spectroscopy: SLSNe-II}

\subsubsection{SLSNe-IIn} 
The spectroscopic evolution of samples of SLSNe-IIn has been poorly explored so far. \cite{2008ApJ...686..467S} and \cite{2010ApJ...709..856S} presented spectroscopic sequences extending to late phases of the luminous SLSNe-IIn SN 2006gy and SN 2006tf, respectively. Very roughly, the observations trace a spectral evolution beginning with blue spectra with a few emission lines with typically simple emission profiles, evolve into more complex spectra with significant structure in the continuum shape and emission line profiles, and eventually an evolution at late times (several hundred days) to spectra with very little continuum dominated by residual h$\alpha$ emission.
There is significant variation among objects so that a more definitive description will have to wait for the publication of larger and better understood samples.

\subsubsection{SLSNe-II} 
There are only a handful of events in this class of SLSNe-II without narrow lines (\citealt{2009ApJ...690.1313G}, \citealt{2009ApJ...690.1303M}, \citealt{2014MNRAS.441..289B}, \citealt{2018MNRAS.475.1046I}). However, their spectral evolution seems to be quite similar: initially the spectra show a hot blue continuum with weak or no features. This is followed (around 20\,d or so after peak) by a significant cooling of the spectral continuum, along with development of more prominent lines, most notable a broad hydrogen H$\alpha$ that grows stronger for several months. As these events rapidly decline in flux, spectroscopic follow-up is usually limited in duration, but the few late-time spectra \citep{2014MNRAS.441..289B,2018MNRAS.475.1046I,2018arXiv180707859B} show only H$\alpha$ emission at $>200$\,d. 

\subsection{UV spectroscopy}
\label{subsec_UV}

\begin{figure}[h]
\includegraphics[width=5in]{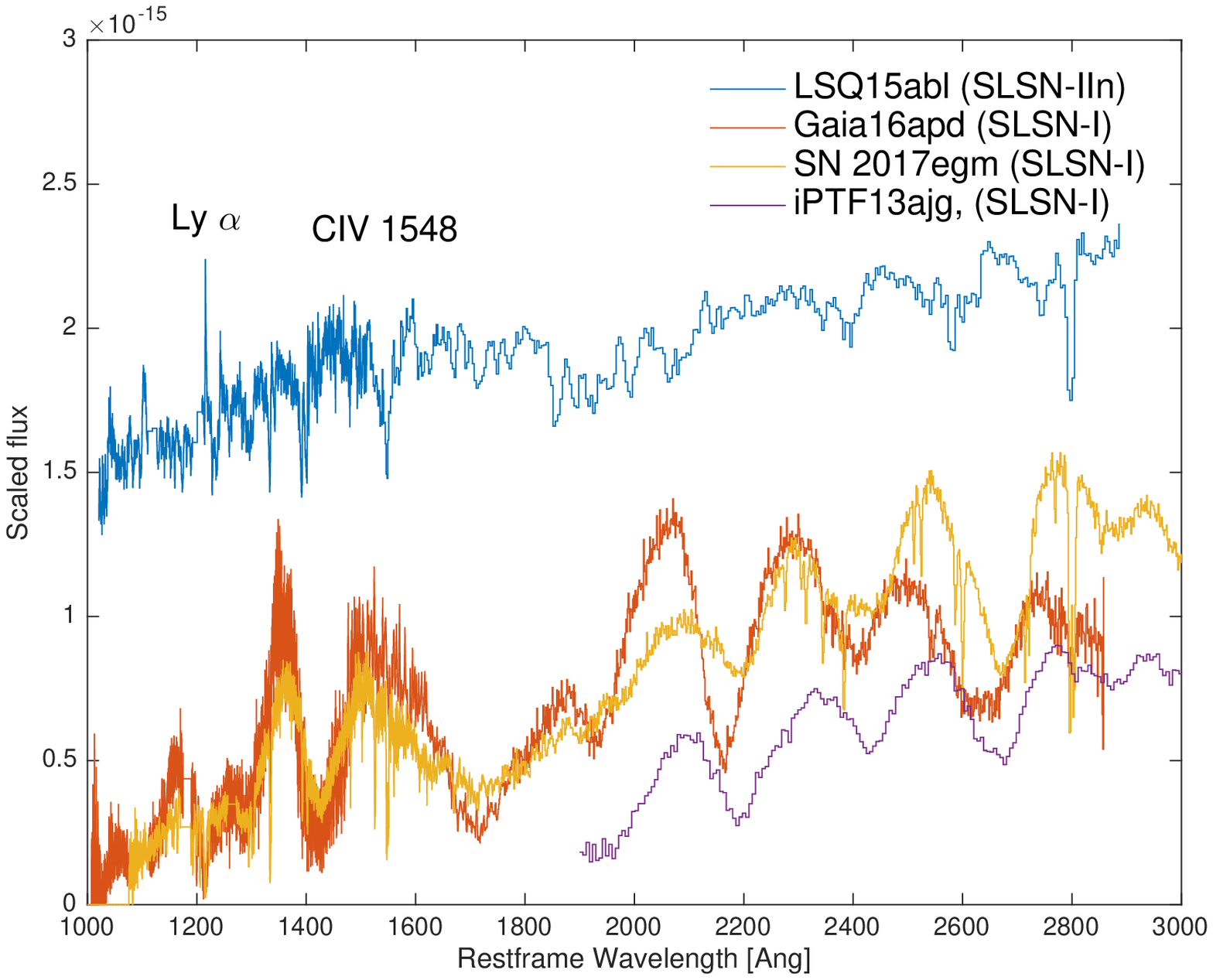}
\caption{Far UV Spectra of SLSNe around peak. SLSNe-I (bottom curves) show a continuum with strong and wide absorption features. The wavelengths and equivalent width of these features are quite homogeneous among the few objects with available observations \citep{2018ApJ...858...91Y}; a ground-based observation of iPTF13ajg \citep{2014ApJ...797...24V} is also shown. The only far UV spectrum of a SLSN-IIn (LSQ15abl) has weaker features and some of these might be blended with host galaxy absorption features; however Ly$\alpha$ emission and CIV 1548\AA\ absorption are broad and are therefore likely associated with the SN \citep{2018arXiv180108241C}.
}
\label{Fig_spec_UV}
\end{figure}

The high UV luminosity of SLSNe made these objects attractive targets for UV spectroscopy using {\it HST}, extending to short wavelengths. Fig.~\ref{Fig_spec_UV} shows available spectroscopy down to 1000\AA\ for SLSNe-I and SLSNe-II \citep{2017ApJ...840...57Y,2018ApJ...858...91Y,2018arXiv180108241C}.

\cite{2018ApJ...858...91Y} compare the few available far UV spectra of SLSNe-I during the hot photospheric phase ($\S$~\ref{subsec_phot_hot}), and find these are quite homogeneous (Fig.~\ref{Fig_spec_UV}). A significant fraction of the total luminosity of these events is emitted in the UV (e.g., $50\%$ below 2500\AA\ for Gaia16apd, \citealt{2017ApJ...840...57Y}) but that the optical-UV flux is not well described by a single BB function, and invoke varying degrees of additional UV flux suppression, perhaps reflecting the metal contents in the outer ejecta. \cite{2018ApJ...854...37S}
collect a sample of high-redshift SLSNe-I with restframe UV spectra (redshifted into the visible) and provide further evidence that these objects have rather homogeneous spectra, peaking around 3000\AA\ with varying UV suppression and marked diversity below 2500\AA. They find no correlation between peak photometric magnitude and spectral properties, nor evidence for spectral evolution at least to modest ($z>1$) redshifts. 

Far-UV Spectra of SLSNe-II are scarce. Fig.~\ref{Fig_spec_UV} shows the single {\it HST} spectrum available. This object shows weaker UV absorption features as well as Ly$\alpha$ emission supporting its classification as a SLSN-IIn. A larger sample of data is needed to further characterize the UV emission from this spectral class. 

\subsection{Polarimetry}

\begin{figure}[h]
\includegraphics[width=5in]{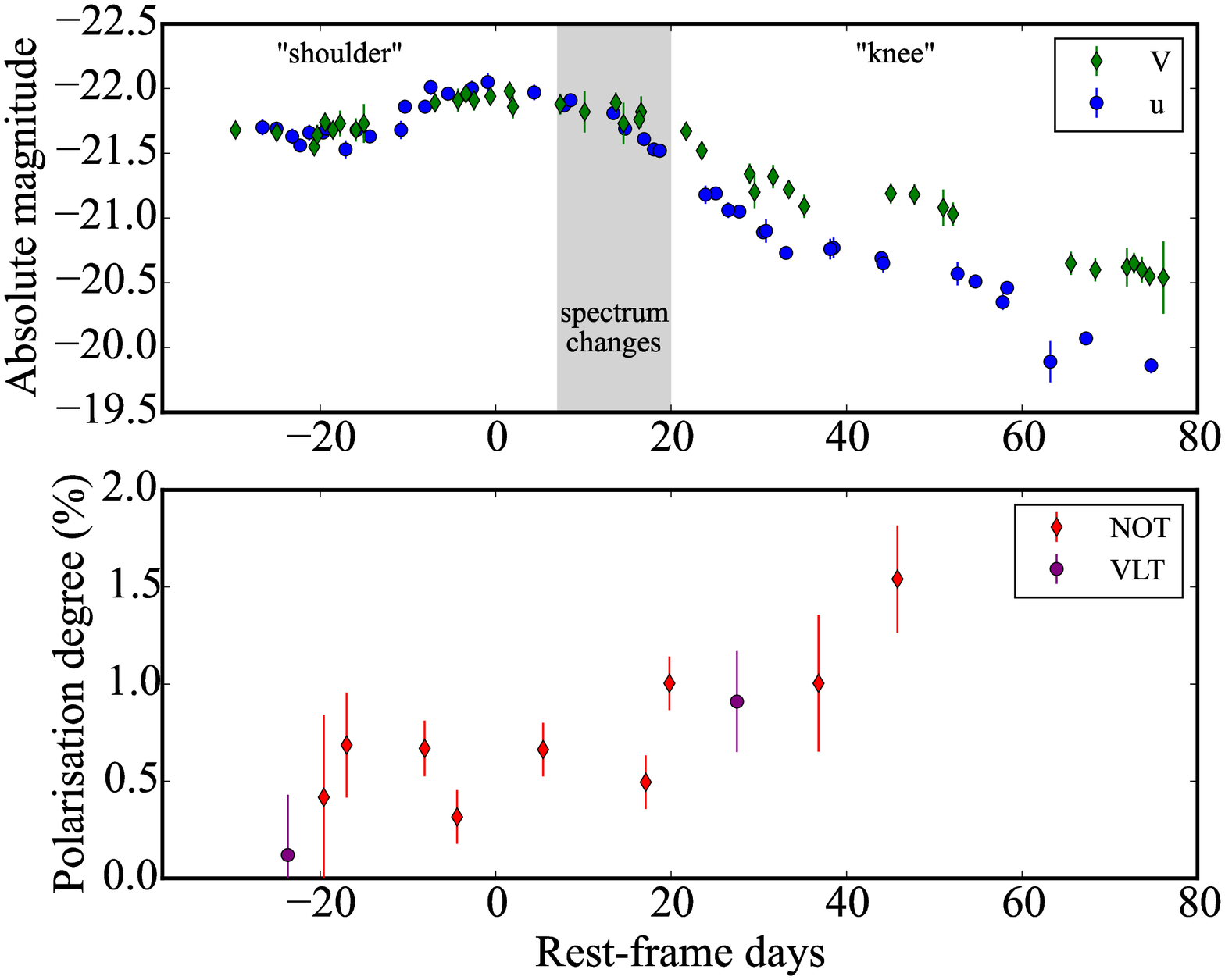}
\caption{Time evolution of the polarisation of SN 2015bn (from \citealt{2017ApJ...837L..14L}; reproduced with permission from the AAS) showing an increase in polarisation following the transition of this object from the hot to the cool photospheric phase.
}
\label{Fig_pol}
\end{figure}

Only a handful of polarimetric studies of SLSNe-I were conducted so far \citep{2015ApJ...815L..10L,2016ApJ...831...79I,2017ApJ...837L..14L,2018ApJ...853...57B}. An interesting result emerged from observations of SN 2015bn, that show a higher level of polarization after peak than prior to peak \citep{2016ApJ...831...79I,2017ApJ...837L..14L}, suggesting an ellipsoidal core \citep{2016ApJ...831...79I} within a more spherical envelope; consistent with analysis by \cite{2018ApJ...853...57B}. The rapid increase in polarization degree that coincides with a sharp spectral transition \citep[Fig~\ref{Fig_pol};][]{2017ApJ...837L..14L} may suggest that the core/envelope border could mark the transition from a relatively pristine C/O progenitor envelope to an inner core of freshly synthesized heavier elements.  

\subsection{Observations at other wavelengths}

\subsubsection{X-rays}

X-ray observations of nearby SLSNe-I using {\it Swift}, Chandra, and XMM-Newton \citep{2013ApJ...763...42O,2013ApJ...771..136L,2017arXiv170405865M} have so far yielded null detections only. \cite{2013ApJ...763...42O} report flux limit in the 0.2-10 keV band around peak magnitude extending from L$_{\rm X}<0.5\times10^{42}$\,erg\,s$^{-1}$ for nearby ($z\approx0.2$) events to about L$_{\rm X}<2\times10^{44}$\,erg\,s$^{-1}$ for less well-observed or higher redshift events. \cite{2013ApJ...771..136L}, \cite{2017MNRAS.468.4642I} and \cite{2017arXiv170405865M} extend this work to a larger sample of events and later time, placing similar limits from pre-peak to $>1000$\,d post-peak for some events. \cite{2017arXiv170405865M} also place much deeper limits (L$_{\rm X}<2\times10^{40}$\,erg\,s$^{-1}$) on emission from PTF12dam prior to peak, interpreting the emission detected in the SN location in deep Chandra observations as arising from the vigorously star-forming host. 

A remarkable exception is the distant (z=1.189) SLSN-I SCP06F6.  
\cite{2013ApJ...771..136L} use XMM observations obtained about 100 restframe days after discovery to measure L$_{\rm X}\approx10^{45}$\,erg\,s$^{-1}$ in 0.2-2 keV X-rays, making SCP06F6 the most luminous X-ray supernova ever detected. Such an emission would have been easily detected for the tens of lower-redshift events monitored with {\it Swift} indicating that either this was an exceptional SLSN-I, or that X-ray emission from SLSNe-I is extremely short-lived and has been missed for most events and fortuitously observed for SCP06F6, or that perhaps the XMM detection is unrelated to the SLSN. Additional similar detections for other SLSNe-I may have significant implications for the physics of these events \citep{2013ApJ...771..136L}.    

\subsubsection{Radio}

GHz radio observations of SLSNe-I have yielded only non-detections so far. \cite{2016ApJ...826...39N} report a deep VLA upper limit 238\,d after maximum. \cite{2018ApJ...853...57B} report deep limits 
(L$_{\rm 10GHz}<5.4\times10^{26}$\,erg\,s$^{-1}\,Hz$$^{-1}$) at the location of the nearby SN2017egm observed around peak, the tightest limits placed so far. These limits rule out an association of these SLSNe-I with GRB-like engines, as even low-luminosity GRB radio afterglow emission would have been recovered \citep{2018ApJ...853...57B}. A radio luminosity similar to those of some regular SNe Ib/c is not ruled out.

\section{PHYSICS OF SUPERLUMINOUS SUPERNOVAE}

Reviewing all the theoretical aspects related to the physics of SLSNe lies far outside the scope of this review. We therefore focus on aspects of SLSN physics that can be related relatively directly to the observations reviewed above. 

\subsection{Energy Source - SLSNe I}
\label{subsec_energy}
A main puzzle about the nature of SLSNe-I is the energy source that powers these very luminous and often very long-lived events. Several possible mechanisms have been proposed and are briefly reviewed here. \cite{2016ApJ...820L..38S} review the maximal theoretical energy output of such models, and find that current observations are so far within the theoretically allowed range. A summary of the models strengths and weaknesses is presented in Table~\ref{SLSN-I_energy}.

\subsubsection{Central engines - magnetars}
\label{subsubsec_magnetars}

A rapidly spinning newborn megnetar was first invoked as an extra power source for a supernova by \cite{2007ApJ...666.1069M}. This idea was further developed and applied to SLSNe by \cite{2010ApJ...719L.204W} and \cite{2010ApJ...717..245K}.
A popular implementation of the model \citep{2013ApJ...770..128I} fits the observed light curves and constrains three physical parameters: the ejecta mass M$_{\rm ej}$, the initial magnetic field B and the initial magnetar period P. Variants of this model have now been fit to large samples of SLSNe-I \citep[e.g.,][]{2017ApJ...842...26L,2017ApJ...850...55N,2018ApJ...860..100D}. \cite{2017ApJ...850...55N} find typical derived values for the physical parameters are B$\approx10^{14}$\,G, P$\approx2$\,ms, and M$_{\rm ej}\approx$5\,M$_{\odot}$, with the mass estimate having a scatter of at least $\times2$ with more massive ejecta typically found for the slowly-declining SLSNe-I; \cite{2018ApJ...860..100D} find similar values for P and somewhat higher values for B. Overall, photometric observations of most SLSNe-I around peak magnitude are well fit by this model, though see \cite{2018ApJ...860..100D} for a few exceptions. 

\cite{2012MNRAS.426L..76D} and \cite{2016MNRAS.458.3455M} calculate synthetic photospheric spectra of SLSNe-I using assumptions motivated by magnetar models and provide good fits to observed spectra. \cite{2017ApJ...835...13J} carry out similar successful modelling of nebular spectra of some slowly-declining SLSNe-I. Overall, magnetar-based models seem to be able to fit both the basic light curves and spectra of SLSNe-I. 

The magnetar model does face some challenges though. As pointed out by \cite{2015ApJ...807L..18N}, the early bump seen in many SLSNe-I before peak ($\S$~\ref{early_bumps}), if interpreted in the context of standard shock-cooling models, requires a large explosion energy ($>10^{52}$\,erg) that might be challenging for neutron-star forming magnetar models. However, \cite{2016ApJ...821...36K} introduce the idea of magnetar shock breakout, that could explain these early bumps.    

Another challenge for this model is that light curve models calculated to fit the early-time observations of SLSNe-I tend to over-predict the emission at late phases (e.g., \citealt{2015MNRAS.452.1567C}, \citealt{2017ApJ...835...58V}).  
To alleviate this discrepancy, it is often assumed that the ejecta become progressively transparent with time, such that the energy deposition from the slowing-down magnetar is not fully trapped at late epochs. Since the nature of the energy output from the magnetar engine and its interaction with the ejecta is unspecified, this may be equivalent to introducing a free time-dependent escape function, making comparison to observations less constraining for the model. 

As the magnetar model predicts a smooth, monotonically decreasing 
energy generation, it cannot naturally explain the post-peak late bumps ($\S$~\ref{late_bumps}) observed in several SLSNe. \cite{2014MNRAS.437..703M} proposed that emerging ionization fronts can create such features, and \cite{2016ApJ...826...39N} suggest a similar idea based on oxygen recombination; however, \cite{2017MNRAS.468.4642I} argue why both of these ideas are unlikely to explain the multiple undulations seen in some slowly-decaying SLSNe-I. Combined with strong evidence for late-time interaction seen in some objects (see below), these authors suggest magnetar-only models are insufficient and hybrid models (e.g., magnetar + interaction) may be required. 

Finally, a consistency issue with this model manifesting as a ``missing mass'' problem appears to arise when comparing ejecta masses derived using the latest data compilation of \cite{2017ApJ...850...55N} with ejecta masses derived from extensive nebular spectroscopy modelling by \cite{2017ApJ...835...13J}. For slowly declining SLSNe-I and in particular for SN 2007bi, the magnetar model light-curve-derived mass ($3.8$\,M$_{\odot}$) is $>4$ times smaller than the lower limit from \cite{2017ApJ...835...13J} ($>15$\,M$_{\odot}$) and inconsistent with previous estimates from \cite{2015ApJ...807L..18N}. However, see \cite{2018arXiv180800510N} for a discussion of the caveats of such a comparison. 

\subsubsection{Central engines - BH accretion}
\label{subsubsec_bh}

\cite{2013ApJ...772...30D} propose fallback accretion onto a newly formed black hole (BH) following the core-collapse of a massive star can power SN-like transients, including SLSNe of both type I and II. This model is similar in many respects to the magnetar model described above, having a central engine energizing the expanding SN ejecta, but unlike the magnetar model, no simple prescription connecting the observed light curve to the model physics exists, beyond the prediction that the late-time energy supply rate falls with time as $t^{-5/3}$.  

While there are few direct comparisons of SLSN observations with predictions based on BH fallback accretion models (beyond those provided by \citealt{2013ApJ...772...30D}), these models do have some potential benefits. In particular, unstable mass accretion onto the BH (as is often the case in, for example, AGN) can naturally explain light curve undulations on various timescales ($\S$~\ref{late_bumps}), and \cite{2016ApJ...821...36K} claim that BH fallback accretion events can also drive a shock breakout flare that can explain the initial bumps observed ($\S$~\ref{early_bumps}). 

With a large reservoir of fallback mass to draw from, and exploiting the high mass to energy conversion possible for thin accretion disks around BHs, the energy budget of BH accretion models is substantially larger than available for NS-forming models (including magnetar models), making them attractive for very energetic events, especially in terms of large integrated luminosity.  

\subsubsection{Radioactivity}
\label{subsubsec_Ni}

Gradual energy injection by radioactive decay of newly synthesized radioactive nuclides (most notably $^{56}$Ni) is generally assumed to be the power source of standard SNe I (both SNe Ia and SNe Ib/c). Some amount of $^{56}$Ni is assumed to form in most variants of massive star explosion models. A particular theoretical variant, predicted to occur for extremely massive stellar cores, involves a rapid contraction of the core following loss of pressure due to electron-positron pair production, leading to a thermonuclear explosion that disrupts the star \citep{1967PhRvL..18..379B,1967ApJ...148..803R}. Such Pair-Instability SN (PISN) explosions are predicted to produce, for high-mass cores, very large amounts of $^{56}$Ni (several solar masses) and therefore result in very luminous SNe \citep{2011ApJ...734..102K}.  

The intrinsic nature of $^{56}$Ni radioactivity as an energy source relates the peak luminosity and integrated energy (both roughly proportional to the total $^{56}$Ni mass) to the ejecta mass and opacity (since Ni and other Fe-group elements have high opacity). One would expect therefore high peak luminosity to be accompanied by long diffusion times for radiation in the ejecta, that should manifest as long rise times to peak. In addition, the energy production rate is known (for $^{56}$Ni and its progeny, $^{56}$Co, the predicted deline rate at late times is $\sim0.01$\,mag\,d$^{-1}$), up to an escape function that is also amenable to calculation, since the energy is deposited into the ejecta by $\gamma$-rays and positrons, and one can calculate the relevant transport equations. This makes radioactivity-based models highly predictive and testable. For PISNe in particular, large $^{56}$Ni masses are produced \citep{2002ApJ...567..532H} from explosions of very massive cores, predicting ejected masses of order $\sim100$\,M$_{\odot}$, and SN rise times of $>100$\,d \citep{2011ApJ...734..102K}.  

In the context of SLSNe-I, the discussion of radioactive models has initially forked into two channels. For some slowly-declining events, in particular for SN 2007bi \citep{2009Natur.462..624G}, PISN models seemed to provide a good fit to available observations. For the group of more rapidly-declining SLSNe-I, radioactive models were disfavored \citep{2011Natur.474..487Q}. This led \cite{2012Sci...337..927G} to propose that there are two separate groups of SLSNe-I, those powered by radioactivity (SLSNe-R) and those
that are not (SLSNe-I). However, additional work challenged the association of slowly-declining SLSNe-I with PISNe \citep{2013Natur.502..346N}. In particular, PISN spectroscopic models seem to disagree with observations for both the photospheric \citep{2012MNRAS.426L..76D} and nebular \citep{2017ApJ...835...13J} phases. In addition, as detailed above, additional analysis does not clearly support a division of SLSNe-I into two separate subclasses based on their light curve shapes.   

PISN models producing the large amounts of $^{56}$Ni relevant for SLSNe-I require very massive exploding cores (with masses $\sim100$\,M$_{\odot}$, e.g., \citealt{2009Natur.462..624G,2011ApJ...734..102K}). Assuming a monotonic initial mass function extending to these extremely massive stars, the detection of even a handful of such objects implies there should be a much larger population of fainter PISN explosions (producing much less $^{56}$Ni from slightly less massive cores; \citealt{2002ApJ...567..532H}). \cite{2018MNRAS.479.3106K} present a plausible candidate for such an event, and the discovery and study of additional similar events by the new generation of massive all-sky transient surveys is an interesting prospect for the nearby future.

At this time, it seems like the radioactive (and in particular, PISN) model is relevant only for a very small subclass of SLSNe-I, that show both a very long rise time, and a slow decline - the two best examples are PS1-14bj \citep{2016ApJ...831..144L} and PTF10nmn \citep{2015MNRAS.454.4357K,2018ApJ...860..100D}. While 
observations of events with spectra similar to the models of \cite{2012MNRAS.426L..76D} have not been published, alternative models \citep{2017MNRAS.464.2854K} seem to be, at least roughly, compatible with observations of some SLSNe-I, making the viability of PISN models more a theoretical than an observational debate. 

\subsubsection{CSM Interaction}
\label{subsubsec_csm}

Interaction with CSM around an exploding star is an efficient mechanism to convert the ejecta kinetic energy into radiation via strong shocks. Such a CSM (see \citealt{2014ARA&A..52..487S} for a review) may result from thick stellar winds, binary interaction and mergers, or stellar eruptions (e.g., of LBV stars). A particular physical mechanism that is often mentioned is pulsational pair-instability \citep[e.g.,][and references therein]{2007Natur.450..390W,2008ApJ...685.1103W,2017ApJ...836..244W}, occurring when massive stellar cores become pair-unstable, but the resulting contraction is halted prior to full core disruption, leading to an ejection of a limited amount of mass - the process may repeat several times prior to final core-collapse.

CSM interaction is likely responsible for powering hydrogen rich type IIn SNe, helium-rich SNe Ibn, rare luminous SN Ia explosions (Ia-CSM) and even some SNe Ic (see \citealt{2017hsn..book..195G} for a recent review of these SN classes). While most interacting SNe show strong emission lines, making this mechanism mostly relevant to SLSNe-II (see below), evidence for interaction with hydrogen-free material largely without prominent emission lines (e.g., \citealt{2014ApJ...785...37B}) motivates consideration of this powering mechanism also for SLSNe-I. 

Several lines of evidence suggest CSM interaction may be important in powering at least some SLSNe-I. Late-time observations of some SLSNe show emerging emission lines of hydrogen \citep[][$\S$~\ref{late_H}]{2017ApJ...848....6Y} or other elements \citep{2018arXiv180804887L} suggesting CSM shells surround the progenitors of some SLSNe-I. \cite{2018arXiv180804887L} show evidence that the PPISN mechanism may have been responsible for the SLSN-I iPTF16eh.

The observation of post-peak undulations in several slowly-declining SLSNe-I can also be explained by interaction of the expanding ejecta with a structured CSM distribution, leading \cite{2017MNRAS.468.4642I} to propose that CSM interaction likely contributes to powering these events. 

\cite{2012ApJ...746..121C} provide a semi-analytic formula to simulate CSM-powered SN light curves, that has been fit to numerous observed SLSNe. While this model is simple and has a lot of freedom (since it depends on the physical properties of both the SN and the CSM) it typically provides fits to light curves that are as good as those provided by Magnetar models \citep[e.g.,][]{2016ApJ...826...39N,2017ApJ...835...58V}.

A major shortcoming of the CSM models is that there are currently no published models that fit observed SLSN-I spectra assuming pure CSM interaction power. \cite{2016MNRAS.457L..79N} point out that having ubiquitous early bumps with similar properties in SLSN-I light curves would be surprising for interaction-powered models, since shock breakout is expected to occur in widely varying locations within complex CSM. The observations of \cite{2017ApJ...835...58V} suggest early bumps may indeed be more diverse than initially thought, as do recent observations of SN 2018bsz \citep{2018arXiv180610609A} and the set of SLSNe-I from DES (C. Inserra, private communication), so early bump properties are probably not a significant challenge for CSM models.     

\subsubsection{Hybrid models}

Hybrid models combine two or more of the specific models discussed above. Such combinations are in some sense expected and perhaps inevitable - for example, megnetar-forming SN explosions may also synthesize some amount of $^{56}$Ni, some late-time CSM interaction may be expected for many massive-star explosions, and so on. Some hybrid models combine a leading model to explain most of the SLSN emission, with a secondary model to explain specific features that are hard to reproduce with the main model - for example, \cite{2017MNRAS.468.4642I} advocate CSM interaction on top of a central engine explosion to explain slowly-declining SLSNe-I and their ubiquitous light curve undulations, while \cite{2017A&A...602A...9C} invoke CSM interaction to explain evidence for an additional luminosity source near peak. The inherent additional parametric flexibility in hybrid models makes them philosophically weaker compared to any single model, but
with accumulated observations, perhaps no single model can explain all data. 

\begin{table}[h]
\tabcolsep7.5pt
\caption{SLSN-I energy source models}
\label{SLSN-I_energy}
\begin{center}
\begin{tabular}{@{}|l|l|l|@{}}
\hline
Model & Strengths & Challenges / {\bf Weaknesses}\\
\hline
Magnetar & Good fits to observed light curves & Early bumps\\
         & Models fit photospheric spectra   & Late decay slope \\
         & Models fit nebular spectra   & {\bf Post-peak bumps} \\
         &                              & {\bf Interaction signatures} \\
         &                              & ``Missing mass'' problem \\
\hline
Black Hole & Naturally explain LC undulations& {\bf Rarely tested with observations}\\
         & Potentially large energy supply & {\bf Interaction signatures} \\
\hline
Radioactivity$^{a}$ & Predictive models fit some events & Spectroscopic behavior debated\\
\hline
CSM Interaction & Interaction signatures in some events& Early bumps\\
                & Late bumps& No predictive spectral models\\
\hline
Hybrid models& Can fit most observations & Hard to test\\
\hline
\end{tabular}
\end{center}
\begin{tabnote}
$^{\rm a}$Only for the most slowly-rising SLSNe-I; 
\end{tabnote}
\end{table}

\subsection{Energy sources - SLSNe-II}

Since most SLSN-II show signs of CSM interactions similar to those of lower luminosity SNe IIn, it is natural to assume that these events are powered by continued conversion of the SN ejecta kinetic energy into radiation by distributions of massive CSM. Some cases require large explosion kinetic energies or large CSM masses, or both (e.g., \citealt{2010ApJ...709..856S}), but no clear cases have been published where an energetic core-collapse explosion of a massive star embedded in a thick CSM cannot suffice. Lacking published samples of events, information about typical physical properties of SLSNe-II are not yet available.  

\subsection{Progenitor stars}
\label{subsec_progenitors}

The association of SLSNe of all types with star-forming hosts (see $\S$~\ref{sec_hosts} below) strongly suggest these are explosions of massive stars. Analysis of host galaxy samples suggests the progenitors of SLSNe-I have a strong preference for sub-solar metallicity ($<0.5$ solar; see $\S$~\ref{sec_hosts} below). 

\subsubsection{Evidence for massive-star progenitors}
Assuming SLSNe are massive star explosions, it is interesting to consider whether they arise from the same mass range as lower-luminosity events (e.g., $8-17$\,M$_{\odot}$ for common SNe II, \citealt{2015PASA...32...16S}), or do they arise from even more massive stars, and whether SLSNe-I and SLSNe-II have similar mass progenitors.

\paragraph{Slowly-rising SLSNe-I} There is strong evidence that at least some SLSNe arise from extremely massive stars. While initial claims for very massive progenitors for SLSNe-I with the most extended light curves,
based on association of some events with PISN models \citep{2009Natur.462..624G} have been challenged (see $\S$~\ref{subsubsec_Ni}), the combined evidence from multiple lines of evidence, including both light curve analysis \citep{2015MNRAS.452.3869N,2016ApJ...831..144L} as well as nebular spectroscopy, suggests very massive progenitors for these events, with ejecta masses of $>20$\,M$_{\odot}$ and derived initial masses M$>40$\,M$_{\odot}$ \citep{2017ApJ...835...13J}.

\paragraph{PPISNe} \cite{2018arXiv180804887L} present evidence, based on spectroscopic analysis of the evolution of transient emission lines, interpreted as echoes from a CSM shell, associating the SLSN-I iPTF16eh with PPISN models. These models can potentially explain the origin of the CSM shell, and indicate an extremely massive progenitor with initial mass $>100$\,M$_{\odot}$. 

\paragraph{SLSNe-IIn} For several SLSNe-IIn with long duration light curves, the long diffusion time and large integrated luminosity require both massive ejecta and a massive CSM shell. Taken together, these require very massive progenitors, perhaps similar to massive LBVs \citep[e.g,][]{2010ApJ...709..856S}. An association with massive LBV-like stars would also explain possible precursor eruptions that may be associated with at least some SLSNe-IIn (see $\S$~\ref{slsn_II_precursors}; \citealt{2014ApJ...789..104O}).

\paragraph{Host galaxy eviedence}

\cite{2015MNRAS.449..917L} and \cite{2018MNRAS.473.1258S} present evidence from the association of SLSNe-I with extreme emission line galaxies (EELGs) suggestion very young ages, and therefore very high progenitor initial masses (M$>60$\,M$_{\odot}$), with some slowly-evolving events (e.g., PTF12dam, \citealt{2015MNRAS.451L..65T}) coming from even higher masses (M$\approx$120\,M$_{\odot}$; see $\S$~\ref{sec_hosts} below). 

\subsubsection{Metallicity}

As shown in $\S$~\ref{sec_rates} below, there is strong evidence from host galaxy population studies that the progenitors of SLSNe-I have low metallicity compared to that of the average massive star population at relatively low redshift (i.e., to recover the observed distribution of host galaxy properties, one must assume a suppression of SLSN-I formation above a metallicity of Z$=0.5$\,Z$_{\odot}$). The lower their natal metallicity, stars of a given initial mass undergo weaker mass loss during their evolution \citep{2012ARA&A..50..107L}. Such stars would thus evolve to have pre-explosion cores that are both more massive and retain more angular momentum, all other parameters being equal. Stars with rapidly rotating core are natural candidates to form central engines such as magnetars ($\S$~\ref{subsubsec_magnetars}) and accreting black holes ($\S$~\ref{subsubsec_bh}). The positive correlation shown by \cite{2015arXiv150602655K} between progenitor luminosity, and hence progenitor mass, and the resulting amount of radioactive $^{56}$Ni synthesized in the resulting explosion, if it extends to large Ni masses, also makes low-metallicity stars favored progenitors for Ni-powered SLSN models ($\S$~\ref{subsubsec_Ni}), and in particular for PISN models for slowly-declining SLSNe-I \citep{2007A&A...475L..19L,2012ARA&A..50..107L,2017MNRAS.464.2854K}.

\subsubsection{Possible connection with Gamma-Ray Bursts}
The association of both SLSNe-I and long-duration GRBs with dwarf, star-forming galaxies, begs the question of a possible connection between these energetic cosmic explosion classes. 
As discussed in $\S$~\ref{sec_hosts} below, while the host galaxies of SLSNe-I and long-duration GRBs show some similarities, they also show significant differences. 

\cite{2015Natur.523..189G} show a possible association between an ultra-long GRB and a supernova that is similar to members of the SLSN-I class. The implications of this discovery are unclear though, as the GRB is a member of a rare, unusual subclass (ultra-long duration GRBs; \citealt{2014ApJ...781...13L}) and the associated supernova is somewhat fainter and has spectral peculiarities (it is UV-faint) compared to other SLSNe-I. In any case, limits from observations of nearby GRBs indicate that most GRBs cannot be associated with SLSN-I explosions - all detected SNe associated with GRBs are fainter than SLSNe-I.  

\section{HOST GALAXIES OF SUPERLUMINOUS SUPERNOVAE}
\label{sec_hosts}

\subsection{Population Studies}

\cite{2011ApJ...727...15N} carried our the first sample study of SLSN hosts, including 17 SLSNe-I and SLSNe-II. Based on the UV and r-band properties of this initial literature compilation, they found that SLSNe occur in blue star-forming galaxies, prefer galaxies with low mass and high specific star formation, and suggested that SLSNe occurrence in high-mass galaxies is suppressed due to high metallicity. 

Additional studies followed using larger samples of SLSNe-I, extensive photometric observations from IR to UV as well as spectroscopy \citep{2014ApJ...787..138L,2015MNRAS.449..917L,2016MNRAS.458...84A,2016ApJ...830...13P,2018MNRAS.473.1258S}. Typical mean properties of SLSN host galaxies are M$<10^9$\,M$_{\odot}$, specific star formation rates of $\sim2$\,Gy$^{-1}$ and low metallicities of 0.5 solar or below, with some exceptions. High resolution {\it HST} imaging \citep{2015ApJ...804...90L,2016MNRAS.458...84A} suggested a dominant compact irregular morphology for these dwarf hosts.    

\cite{2014ApJ...787..138L} suggested SLSNe-I and long-duration GRBs occur in similar host galaxies within the overall population of star-forming galaxies, but this result was not supported by later studies \citep{2015MNRAS.449..917L,2016MNRAS.458...84A,2018MNRAS.473.1258S}. 

Taken together, the most recent results suggest that SLSNe-I likely occur mostly in compact, dwarf irregular galaxies, often with extremely strong emission lines \citep[EELGs; EW$_{\rm OIII}>100$\AA,][]{2015MNRAS.449..917L}, with an effective metallicity suppression above $0.5$ solar \citep{2016ApJ...830...13P, 2017MNRAS.470.3566C, 2018MNRAS.473.1258S}. The hosts of SLSNe-IIn are different \citep{2015MNRAS.449..917L}, span a very wide range of luminosities, masses and metallicities \citep{2016MNRAS.458...84A}, possibly with metallicity suppression above 0.8 solar \citep{2018MNRAS.473.1258S} (though see \citealt{2016ApJ...830...13P}) but are distinct from host galaxies of the general core-collapse SN population \citep{2018MNRAS.473.1258S}, for example in the fraction of events exploding in very low luminosity galaxies \citep{2016MNRAS.458...84A}.

\subsection{Absorption Spectroscopy}

The high UV luminosity of SLSNe-I makes them attractive targets to UV absorption line studies that probe the ISM in and around the host galaxies of these events. Ground-based studies of this sort can be carried out for events with redshift $z>0.5$ once important UV metal lines are redshifted below the atmospheric cutoff. \cite{2012ApJ...755L..29B} demonstrated this for a high-redshift SLSN-I from PS1. \cite{2014ApJ...797...24V} carried out an absorption line analysis for the high-redshift PTF object iPTF13ajg. Comparing Mg line absorption strengths for a sample of SLSNe-I and long-duration GRBs, they show that also in this respect, SLSNe-I hosts are distinct from GRB host galaxies in having weaker absorptions.

\section{RATES}
\label{sec_rates}

\cite{2013MNRAS.431..912Q} use observations from the TSS survey to estimate the rate of SLSNe at low redshifts ($z\approx0.15$). For SLSNe-I, they find a rate of $35^{+84}_{-29}$ events\,Gpc$^{-3}$\,y$^{-1}$ for a Hubble constant H$=71$\,km\,s$^{-1}$\,Mpc$^{-1}$, based on a single event. The TSS survey discovered 3 SLSNe-II (2 SLSNe-IIn and a single SLSN-II, SN 2008es). The combined rate derived by \cite{2013MNRAS.431..912Q} from these events is $151^{+151}_{-82}$ events\,Gpc$^{-3}$\,y$^{-1}$. Note however that while the small TSS sample includes a single SLSN-II (without narrow lines) out of the sample of three SLSNe-II in total, the fraction of such SLSNe-II in literature compilations is much smaller (of order 10\%; e.g., \citealt{2012Sci...337..927G}).

As noted by \cite{2012Sci...337..927G}, the rates of SLSNe-I at low redshift are substantially lower than the rates of SNe Ib/c (25800 events\,Gpc$^{-3}$\,y$^{-1}$; \citealt{2011MNRAS.412.1473L}) as well as from the rate of low-redshift long-duration GRBs ($380^{+620}_{-225}$ events\,Gpc$^{-3}$\,y$^{-1}$, \citealt{2007ApJ...657L..73G}). If verified by better statistics, the fact that SLSN-I rates fall below that of GRBs suggests that if these events arise from massive progenitors in similar galaxies, the initial mass range of SLSN-I progenitors should be higher than that of GRB progenitors.

\cite{2017MNRAS.464.3568P} measured the rate of SLSNe-I at higher redshifts ($z\approx1.1$) using observations by the SNLS survey, and find a rate of $91^{+76}_{-36}$ events\,Gpc$^{-3}$\,y$^{-1}$. Combined with the measurement at low redshift, the SLSN-I rate appears to rise in a manner consistent with the evolution of the cosmic star formation rate. These authors also find that the SLSN-I rate is substantially below that of SNe Ib/c or long GRBs, but is consistent with the rare class of ultra-long GRBs.   

\cite{2012Natur.491..228C} used the detection of two likely SLSNe (lacking spectral confirmation) at very high redshifts (z=2.04 and z=3.9) to estimate the high-redshift rate of SLSNe, finding a rate of $\approx400$ events\,Gpc$^{-3}$\,y$^{-1}$ with very large errors. If these events are related to lower-luminosity SLSNe-I, then this rate is marginally above that expected from scaling the measurement of \cite{2017MNRAS.464.3568P} by the cosmic star formation rate, suggesting that the fraction of SLSN progenitors of the massive star population may have been larger in the past.   

\section{POSSIBLE USE IN COSMOLOGY}
\label{sec_cosmology}

Assembling a sample of SLSNe-I from the literature and their own observations, \cite{2013MNRAS.431..912Q} noted that the peak magnitude of SLSNe-I tends to cluster quite strongly around the mean value (M$=-21.7$\,mag) and speculated that this class of objects may become useful distance estimators for cosmology.

\cite{2014ApJ...796...87I} investigated this option further using a larger sample of relatively well-observed events, and suggested that correlations between the peak magnitude and the light curve shape, as well as color, could help standardize these objects as cosmological probes, perhaps competitive with SNe Ia. \cite{2018ApJ...860..100D} attempted to conducted a similar analysis on their sample of SLSNe-I, but could not provide further support to this result. \cite{2018ApJ...854..175I} inspected correlations among photometric parameters of SLSNe-I, suggesting these do support their division into two photometric subclasses (slowly- and rapidly-declining events), and that these correlations can further assist in cosmological use of these events. 

\section{FUTURE PROSPECTS AND OPEN QUESTIONS}
\label{sec_future}

The study of SLSNe is a young subject with less than a dozen years having passed since the publications of the first pioneering works. It is clear that significant additional work remains in order to observationally characterize these objects and to better understand their physics. I highlight here several important open questions and areas where additional important progress is likely.

A major open question highlighted in this work is the energy source powering these extremely luminous and long-lived events. Are all SLSNe-I powered by the same source or are there two, or more, sub-groups? Are some or all SLSNe-II similar to SLSNe-I in this sense? Is a central engine definitely required, and if
so, is it a Magnetar, an accreting black hole, or perhaps both channels are at work? Have we ever observed the results of the PISN and PPISN mechanisms? These and additional question remain open, and continue to drive a wide and extensive observational effort. 

Type II SLSNe are significantly less well studied compared to SLSNe-I, and in particular sample studies are missing. Leloudas et al. (in preparation, see $\S$~\ref{subsubsec_II_phot}) should provide at least an initial characterization of SLSNe-IIn in the nearby Universe, but additional studies, in particular also at higher redshifts, are certainly called for. 

With multiple transient surveys such as ASASSN, ATLAS, PS and ZTF monitoring almost the entire sky (both hemispheres, from ground and from space using Gaia) with typical cadence of a few days or less, long-lasting and luminous nearby SLSNe should be detected with high completeness, as demonstrated, e.g., with the discoveries of Gaia16apd \citep{2017ApJ...835L...8N,2017ApJ...840...57Y,2017MNRAS.469.1246K}, SN 2017egm \citep{2018ApJ...853...57B} and SN 2018bsz \citep{2018arXiv180610609A}.
As more and more events are studied at lower and lower distances, new observational windows open to study these luminous events with diverse instrumentation. I am especially excited about the prospects to closely monitor a nearby SLSN-I in X-rays in order to test whether an X-ray burst similar to the one detected by \cite{2013ApJ...771..136L} can be recovered; such an event could provide strong support to central-engine models.

Moving to the other extreme, the high UV luminosity of SLSNe makes them perhaps the best targets for studies of transients at high redshift. Following the pioneering work of \cite{2012Natur.491..228C}, several works have already highlighted the prospects for detecting SLSNe at very high redshift using, for example, LSST, as well as future space missions such as Euclid \citep{2018A&A...609A..83I} and WFIRST \citep{2018ApJ...854...37S}. \cite{2017arXiv171007005W} show that JWST could conduct a powerful survey for extremely high-redshift transients, and that $z>10$ SLSNe-I could be detected by JWST in the NIR bands; in fact, SLSNe (if they occur at very high redshifts) may be the most luminous source of light in the early universe, easily outshining infant galaxies. SLSNe at high redshift could be a focus of attention in the coming decade not only for the massive star, supernova and time-domain astronomy community, but also for those interested in the first stars, the early universe and the first sources of ionizing radiation.   

\section{SUMMARY}

This review attempts to summarize our knowledge about SLSNe accumulated mostly in the last decade or so, from an observational perspective. SLSNe are spectroscopically 
classified into hydrogen-poor SLSNe-I and hydrogen-rich SLSNe-II. 

Hydrogen-poor SLSNe-I are quite well characterized. This class includes events with a peak magnitude above a threshold of M$_{g}=-19.8$\,mag or so, based on a combination of photometric and spectroscopic studies of low-redshift SLSN-I samples ($\S$~\ref{subsec_thresh}). The light curves evolve more slowly than those of lower-luminosity SNe Ib/c, with typical rise times $>20$\,d and a decay time distribution that extends to very slowly-declining events (Fig.~\ref{Fig_LC}). Light curve bumps and undulations are observed both at early ($\S$~\ref{early_bumps}) and late, post-peak ($\S$~\ref{late_bumps}) phases. 

SLSNe-I evolve through several spectroscopic phases. A initial hot photospheric phase, charaterised by a blue continuum with high black-body temperature ($>12$\,kK) with absorption features mostly of carbon and oxygen in visible light (Fig.~\ref{Fig_class_slsn_I}) and several strong UV features (Fig.~\ref{Fig_spec_UV}), typically persists until peak magnitude or a few weeks later (Fig.~\ref{Fig_LC}). This is followed by a cool photospheric phase with spectra similar to those of SNe Ic (Fig.~\ref{Fig_class_slsn_I_late}) that further evolve into the nebular phase (Fig.~\ref{Fig_slsn_nebular}) $>150$\,d after peak. 

Hydrogen-rich SLSNe-II are not nearly as well studied. The majority of these events show strong narrow emission lines of hydrogen and are spectroscopically similar to lower-luminosity SNe IIn ($\S$~\ref{Fig_class_slsn_II}); it is not clear if there is a threshold separating the SLSN-IIn population from their more common low-luminosity cousins. The mean peak magnitude of this population is M$_{r}\approx-21.1\pm0.5$\,mag, and its rise and decay times are long, even compared to SLSNe-I ($\S$~\ref{subsubsec_II_phot}). A handful of rare SLSNe-II show broad (but not narrow) hydrogen features, and some spectral similarities to SLSNe-I ($\S$~\ref{subsubsec_slsn_II_I}). A single event (PTF10hgi) should likely be spectroscopically classified as a SLSN-IIb.    

The energy source of SLSNe-I is still an open question, with viable models including central-engine models driven by a newborn rapidly-spinning magnetar or an accreting black hole, interaction with hydrogen-poor CSM, or, perhaps for the most slowly-evolving events, models powered by large amounts of radioactive $^{56}$Ni ($\S$~\ref{subsec_energy}). The energy source for SLSNe-II is even less well understood. It is clear that interaction with CSM plays a major role for most of these events, but whether the underlying explosion is similar to that of lower-luminosity events is not known. The progenitor stars of SLSNe are young, massive stars; there is evidence that at least some events have extremely massive progenitors (M$>40$\,M$_{\odot}$; $\S$~\ref{subsec_progenitors}).

The host galaxies of SLSNe-I are typically compact, irregular dwarf galaxies with low stellar mass and high star formation rate, and often also with very strong emission lines. SLSN-I production seems to be suppressed above a metallicity threshold of Z$=0.5$\,Z$_{\odot}$ ($\S$~\ref{sec_hosts}). The hosts of SLSNe-II span a wider range of masses, but are also distinct from the hosts of lower-luminosity events. The rates of SLSNe are still rather poorly measured due to small number statistics, however, these events are rare, with volumetric rates at least two orders of magnitude below those of normal core-collapse SNe; SLSNe-I are likely rarer also from long-duration GRBs ($\S$~\ref{sec_rates}).

SLSNe-I have been proposed as cosmological probes visible to high redshifts, and several papers investigate this possibility, though the results are not yet conclusive ($\S$~\ref{sec_cosmology}). With likely near-term prospects extending from detection of additional very nearby SLSNe to the most distant events ($\S$~\ref{sec_future}), the prospects of this young field of study for additional rapid development appear bright.   

\section*{DISCLOSURE STATEMENT}
The authors are not aware of any affiliations, memberships, funding, or financial holdings that
might be perceived as affecting the objectivity of this review. 

\section*{ACKNOWLEDGMENTS}
The authors thanks G. Leloudas, S. Schulze, I. Manulis, O. Yaron, P. Vreeswijk, A. Horesh and C. Inserra for help and advice. This research was supported by the EU via ERC grant No. 725161, the Quantum Universe I-Core program, the ISF, the BSF Transformative program and by a Kimmel award.
%





\begin{thebibliography}{00}


\bibitem[Anderson et al.(2018)]{2018arXiv180610609A} Anderson, J.~P., Pessi, P.~J., Dessart, L., et al.\ 2018, arXiv:1806.10609 
\bibitem[Angus et al.(2016)]{2016MNRAS.458...84A} Angus, C.~R., Levan, A.~J., Perley, D.~A., et al.\ 2016, \mnras, 458, 84 
\bibitem[Arcavi et al.(2010)]{2010ApJ...721..777A} Arcavi, I., Gal-Yam, A., Kasliwal, M.~M., et al.\ 2010, \apj, 721, 777 
\bibitem[Arcavi et al.(2017)]{2017Natur.551..210A} Arcavi, I., Howell, D.~A., Kasen, D., et al.\ 2017, \nat, 551, 210 


\bibitem[Barkat et al.(1967)]{1967PhRvL..18..379B} Barkat, Z., Rakavy, G., \& Sack, N.\ 1967, Physical Review Letters, 18, 379 
\bibitem[Ben-Ami et al.(2014)]{2014ApJ...785...37B} Ben-Ami, S., Gal-Yam, A., Mazzali, P.~A., et al.\ 2014, \apj, 785, 37 
\bibitem[Benetti et al.(2014)]{2014MNRAS.441..289B} Benetti, S., Nicholl, M., Cappellaro, E., et al.\ 2014, \mnras, 441, 289
\bibitem[Berger et al.(2012)]{2012ApJ...755L..29B} Berger, E., Chornock, R., Lunnan, R., et al.\ 2012, \apjl, 755, L29 
\bibitem[Bhirombhakdi et al.(2018)]{2018arXiv180707859B} Bhirombhakdi, K., Chornock, R., Miller, A.~A., et al.\ 2018, arXiv:1807.07859 
\bibitem[Bose et al.(2018)]{2018ApJ...853...57B} Bose, S., Dong, S., Pastorello, A., et al.\ 2018, \apj, 853, 57 


\bibitem[Chambers et al.(2016)]{2016arXiv161205560C} Chambers, K.~C., Magnier, E.~A., Metcalfe, N., et al.\ 2016, arXiv:1612.05560 
\bibitem[Chatzopoulos et al.(2011)]{2011ApJ...729..143C} Chatzopoulos, E., Wheeler, J.~C., Vinko, J., et al.\ 2011, \apj, 729, 143 
\bibitem[Chatzopoulos et al.(2012)]{2012ApJ...746..121C} Chatzopoulos, E., Wheeler, J.~C., \& Vinko, J.\ 2012, \apj, 746, 121 
\bibitem[Chen et al.(2015)]{2015MNRAS.452.1567C} Chen, T.-W., Smartt, S.~J., Jerkstrand, A., et al.\ 2015, \mnras, 452, 1567 
\bibitem[Chen et al.(2017)]{2017MNRAS.470.3566C} Chen, T.-W., Smartt, S.~J., Yates, R.~M., et al.\ 2017, \mnras, 470, 3566 
\bibitem[Chen et al.(2017)]{2017A&A...602A...9C} Chen, T.-W., Nicholl, M., Smartt, S.~J., et al.\ 2017, \aap, 602, A9 
\bibitem[Cooke et al.(2012)]{2012Natur.491..228C} Cooke, J., Sullivan, M., Gal-Yam, A., et al.\ 2012, \nat, 491, 228 
\bibitem[Curtin et al.(2018)]{2018arXiv180108241C} Curtin, C., Cooke, J., Moriya, T.~J., et al.\ 2018, arXiv:1801.08241 

\bibitem[De Cia et al.(2018)]{2018ApJ...860..100D} De Cia, A., Gal-Yam, A., Rubin, A., et al.\ 2018, \apj, 860, 100 
\bibitem[Dessart et al.(2012)]{2012MNRAS.426L..76D} Dessart, L., Hillier, D.~J., Waldman, R., Livne, E., \& Blondin, S.\ 2012, \mnras, 426, L76 
\bibitem[Dexter \& Kasen(2013)]{2013ApJ...772...30D} Dexter, J., \& Kasen, D.\ 2013, \apj, 772, 30 
\bibitem[Dong et al.(2016)]{2016Sci...351..257D} Dong, S., Shappee, B.~J., Prieto, J.~L., et al.\ 2016, Science, 351, 257 
\bibitem[Drake et al.(2009)]{2009ApJ...696..870D} Drake, A.~J., Djorgovski, S.~G., Mahabal, A., et al.\ 2009, \apj, 696, 870 

\bibitem[Gal-Yam et al.(2009)]{2009Natur.462..624G} Gal-Yam, A., Mazzali, P., Ofek, E.~O., et al.\ 2009, \nat, 462, 624 
\bibitem[Gal-Yam(2012)]{2012Sci...337..927G} Gal-Yam, A.\ 2012, Science, 337, 927 
\bibitem[Gal-Yam et al.(2013)]{2013PASP..125..749G} Gal-Yam, A., Mazzali, P.~A., Manulis, I., \& Bishop, D.\ 2013, \pasp, 125, 749 
\bibitem[Gal-Yam(2017)]{2017hsn..book..195G} Gal-Yam, A.\ 2017, Handbook of Supernovae, ISBN 978-3-319-21845-8.~Springer International Publishing AG, 2017, p.~195, 195 
\bibitem[Gal-Yam(2018)]{2018arXiv180608224G} Gal-Yam, A.\ 2018, arXiv:1806.08224 
\bibitem[Gezari et al.(2009)]{2009ApJ...690.1313G} Gezari, S., Halpern, J.~P., Grupe, D., et al.\ 2009, \apj, 690, 1313 
\bibitem[Greiner et al.(2015)]{2015Natur.523..189G} Greiner, J., Mazzali, P.~A., Kann, D.~A., et al.\ 2015, \nat, 523, 189 
\bibitem[Guetta \& Della Valle(2007)]{2007ApJ...657L..73G} Guetta, D., \& Della Valle, M.\ 2007, \apjl, 657, L73 

\bibitem[Heger \& Woosley(2002)]{2002ApJ...567..532H} Heger, A., \& Woosley, S.~E.\ 2002, \apj, 567, 532 
\bibitem[Howell et al.(2013)]{2013ApJ...779...98H} Howell, D.~A., Kasen, D., Lidman, C., et al.\ 2013, \apj, 779, 98 
\bibitem[Howell(2017)]{2017hsn..book..431H} Howell, D.~A.\ 2017, Handbook of Supernovae, ISBN 978-3-319-21845-8.~Springer International Publishing AG, 2017, p.~431, 431 


\bibitem[Inserra et al.(2013)]{2013ApJ...770..128I} Inserra, C., Smartt, S.~J., Jerkstrand, A., et al.\ 2013, \apj, 770, 128 
\bibitem[Inserra \& Smartt(2014)]{2014ApJ...796...87I} Inserra, C., \& Smartt, S.~J.\ 2014, \apj, 796, 87 
\bibitem[Inserra et al.(2016)]{2016ApJ...831...79I} Inserra, C., Bulla, M., Sim, S.~A., \& Smartt, S.~J.\ 2016, \apj, 831, 79 
\bibitem[Inserra et al.(2017)]{2017MNRAS.468.4642I} Inserra, C., Nicholl, M., Chen, T.-W., et al.\ 2017, \mnras, 468, 4642 
\bibitem[Inserra et al.(2018)]{2018ApJ...854..175I} Inserra, C., Prajs, S., Gutierrez, C.~P., et al.\ 2018, \apj, 854, 175 
\bibitem[Inserra et al.(2018)]{2018MNRAS.475.1046I} Inserra, C., Smartt, S.~J., Gall, E.~E.~E., et al.\ 2018, \mnras, 475, 1046 
\bibitem[Inserra et al.(2018)]{2018A&A...609A..83I} Inserra, C., Nichol, R.~C., Scovacricchi, D., et al.\ 2018, \aap, 609, A83 


\bibitem[Jerkstrand et al.(2017)]{2017ApJ...835...13J} Jerkstrand, A., Smartt, S.~J., Inserra, C., et al.\ 2017, \apj, 835, 13 


\bibitem[Kangas et al.(2017)]{2017MNRAS.469.1246K} Kangas, T., Blagorodnova, N., Mattila, S., et al.\ 2017, \mnras, 469, 1246 
\bibitem[Kasen \& Bildsten(2010)]{2010ApJ...717..245K} Kasen, D., \& Bildsten, L.\ 2010, \apj, 717, 245 
\bibitem[Kasen et al.(2011)]{2011ApJ...734..102K} Kasen, D., Woosley, S.~E., \& Heger, A.\ 2011, \apj, 734, 102 
\bibitem[Kasen et al.(2016)]{2016ApJ...821...36K} Kasen, D., Metzger, B.~D., \& Bildsten, L.\ 2016, \apj, 821, 36 
\bibitem[Kelly \& Kirshner(2012)]{2012ApJ...759..107K} Kelly, P.~L., \& Kirshner, R.~P.\ 2012, \apj, 759, 107 
\bibitem[Kozyreva \& Blinnikov(2015)]{2015MNRAS.454.4357K} Kozyreva, A., \& Blinnikov, S.\ 2015, \mnras, 454, 4357 
\bibitem[Kozyreva et al.(2017)]{2017MNRAS.464.2854K} Kozyreva, A., Gilmer, M., Hirschi, R., et al.\ 2017, \mnras, 464, 2854 
\bibitem[Kozyreva et al.(2018)]{2018MNRAS.479.3106K} Kozyreva, A., Kromer, M., Noebauer, U.~M., \& Hirschi, R.\ 2018, \mnras, 479, 3106 
\bibitem[Kushnir(2015)]{2015arXiv150602655K} Kushnir, D.\ 2015, arXiv:1506.02655 


\bibitem[Langer et al.(2007)]{2007A&A...475L..19L} Langer, N., Norman, C.~A., de Koter, A., et al.\ 2007, \aap, 475, L19
\bibitem[Langer(2012)]{2012ARA&A..50..107L} Langer, N.\ 2012, \araa, 50, 107 
\bibitem[Law et al.(2009)]{2009PASP..121.1395L} Law, N.~M., Kulkarni, S.~R., Dekany, R.~G., et al.\ 2009, \pasp, 121, 1395 
\bibitem[Leloudas et al.(2012)]{2012A&A...541A.129L} Leloudas, G., Chatzopoulos, E., Dilday, B., et al.\ 2012, \aap, 541, A129 
\bibitem[Leloudas et al.(2015)]{2015ApJ...815L..10L} Leloudas, G., Patat, F., Maund, J.~R., et al.\ 2015, \apjl, 815, L10 
\bibitem[Leloudas et al.(2015)]{2015MNRAS.449..917L} Leloudas, G., Schulze, S., Kr{\"u}hler, T., et al.\ 2015, \mnras, 449, 917 
\bibitem[Leloudas et al.(2016)]{2016NatAs...1E...2L} Leloudas, G., Fraser, M., Stone, N.~C., et al.\ 2016, Nature Astronomy, 1, 0002 
\bibitem[Leloudas et al.(2017)]{2017ApJ...837L..14L} Leloudas, G., Maund, J.~R., Gal-Yam, A., et al.\ 2017, \apjl, 837, L14 
\bibitem[Levan et al.(2013)]{2013ApJ...771..136L} Levan, A.~J., Read, A.~M., Metzger, B.~D., Wheatley, P.~J., \& Tanvir, N.~R.\ 2013, \apj, 771, 136 
\bibitem[Levan et al.(2014)]{2014ApJ...781...13L} Levan, A.~J., Tanvir, N.~R., Starling, R.~L.~C., et al.\ 2014, \apj, 781, 13 
\bibitem[Li et al.(2011)]{2011MNRAS.412.1473L} Li, W., Chornock, R., Leaman, J., et al.\ 2011, \mnras, 412, 1473 
\bibitem[Liu et al.(2017)]{2017ApJ...845...85L} Liu, Y.-Q., Modjaz, M., \& Bianco, F.~B.\ 2017, \apj, 845, 85 
\bibitem[Liu et al.(2017)]{2017ApJ...842...26L} Liu, L.-D., Wang, S.-Q., Wang, L.-J., et al.\ 2017, \apj, 842, 26 
\bibitem[Lunnan et al.(2014)]{2014ApJ...787..138L} Lunnan, R., Chornock, R., Berger, E., et al.\ 2014, \apj, 787, 138 
\bibitem[Lunnan et al.(2015)]{2015ApJ...804...90L} Lunnan, R., Chornock, R., Berger, E., et al.\ 2015, \apj, 804, 90 
\bibitem[Lunnan et al.(2016)]{2016ApJ...831..144L} Lunnan, R., Chornock, R., Berger, E., et al.\ 2016, \apj, 831, 144 
\bibitem[Lunnan et al.(2018)]{2018ApJ...852...81L} Lunnan, R., Chornock, R., Berger, E., et al.\ 2018, \apj, 852, 81 
\bibitem[Lunnan et al.(2018)]{2018arXiv180804887L} Lunnan, R., Fransson, C., Vreeswijk, P.~M., et al.\ 2018, arXiv:1808.04887 

\bibitem[Maeda et al.(2007)]{2007ApJ...666.1069M} Maeda, K., Tanaka, M., Nomoto, K., et al.\ 2007, \apj, 666, 1069 
\bibitem[Margutti et al.(2017)]{2017arXiv170405865M} Margutti, R., Chornock, R., Metzger, B.~D., et al.\ 2017, arXiv:1704.05865 
\bibitem[Mazzali et al.(2016)]{2016MNRAS.458.3455M} Mazzali, P.~A., Sullivan, M., Pian, E., Greiner, J., \& Kann, D.~A.\ 2016, \mnras, 458, 3455 
\bibitem[Metzger et al.(2014)]{2014MNRAS.437..703M} Metzger, B.~D., Vurm, I., Hasco{\"e}t, R., \& Beloborodov, A.~M.\ 2014, \mnras, 437, 703 
\bibitem[Miller et al.(2009)]{2009ApJ...690.1303M} Miller, A.~A., Chornock, R., Perley, D.~A., et al.\ 2009, \apj, 690, 1303 
\bibitem[Miller et al.(2010)]{2010MNRAS.404..305M} Miller, A.~A., Silverman, J.~M., Butler, N.~R., et al.\ 2010, \mnras, 404, 305 
\bibitem[Modjaz et al.(2016)]{2016ApJ...832..108M} Modjaz, M., Liu, Y.~Q., Bianco, F.~B., \& Graur, O.\ 2016, \apj, 832, 108 
\bibitem[Moriya et al.(2018)]{2018SSRv..214...59M} Moriya, T.~J., Sorokina, E.~I., \& Chevalier, R.~A.\ 2018, \ssr, 214, \#59 


\bibitem[Neill et al.(2011)]{2011ApJ...727...15N} Neill, J.~D., Sullivan, M., Gal-Yam, A., et al.\ 2011, \apj, 727, 15 
\bibitem[Nicholl et al.(2013)]{2013Natur.502..346N} Nicholl, M., Smartt, S.~J., Jerkstrand, A., et al.\ 2013, \nat, 502, 346 
\bibitem[Nicholl et al.(2015)]{2015ApJ...807L..18N} Nicholl, M., Smartt, S.~J., Jerkstrand, A., et al.\ 2015, \apjl, 807, L18 
\bibitem[Nicholl et al.(2015)]{2015MNRAS.452.3869N} Nicholl, M., Smartt, S.~J., Jerkstrand, A., et al.\ 2015, \mnras, 452, 3869 
\bibitem[Nicholl \& Smartt(2016)]{2016MNRAS.457L..79N} Nicholl, M., \& Smartt, S.~J.\ 2016, \mnras, 457, L79 
\bibitem[Nicholl et al.(2016)]{2016ApJ...826...39N} Nicholl, M., Berger, E., Smartt, S.~J., et al.\ 2016, \apj, 826, 39 
\bibitem[Nicholl et al.(2016)]{2016ApJ...828L..18N} Nicholl, M., Berger, E., Margutti, R., et al.\ 2016, \apjl, 828, L18 
\bibitem[Nicholl et al.(2017)]{2017ApJ...835L...8N} Nicholl, M., Berger, E., Margutti, R., et al.\ 2017, \apjl, 835, L8 
\bibitem[Nicholl et al.(2017)]{2017ApJ...850...55N} Nicholl, M., Guillochon, J., \& Berger, E.\ 2017, \apj, 850, 55 
\bibitem[Nicholl et al.(2018)]{2018arXiv180800510N} Nicholl, M., Berger, E., Blanchard, P.~K, Gomez, S., \& Chornock, R.\ 2018, arXiv:1808.00510 

\bibitem[Ofek et al.(2007)]{2007ApJ...659L..13O} Ofek, E.~O., Cameron, P.~B., Kasliwal, M.~M., et al.\ 2007, \apjl, 659, L13 
\bibitem[Ofek et al.(2013)]{2013ApJ...763...42O} Ofek, E.~O., Fox, D., Cenko, S.~B., et al.\ 2013, \apj, 763, 42 
\bibitem[Ofek et al.(2014)]{2014ApJ...789..104O} Ofek, E.~O., Sullivan, M., Shaviv, N.~J., et al.\ 2014, \apj, 789, 104 


\bibitem[Pastorello et al.(2010)]{2010ApJ...724L..16P} Pastorello, A., Smartt, S.~J., Botticella, M.~T., et al.\ 2010, \apjl, 724, L16 
\bibitem[Perley et al.(2016)]{2016ApJ...830...13P} Perley, D.~A., Quimby, R.~M., Yan, L., et al.\ 2016, \apj, 830, 13 
\bibitem[Prajs et al.(2017)]{2017MNRAS.464.3568P} Prajs, S., Sullivan, M., Smith, M., et al.\ 2017, \mnras, 464, 3568 
\bibitem[Prentice et al.(2016)]{2016MNRAS.458.2973P} Prentice, S.~J., Mazzali, P.~A., Pian, E., et al.\ 2016, \mnras, 458, 2973 

\bibitem[Quimby et al.(2007)]{2007ApJ...668L..99Q} Quimby, R.~M., Aldering, G., Wheeler, J.~C., et al.\ 2007, \apjl, 668, L99 
\bibitem[Quimby et al.(2011)]{2011Natur.474..487Q} Quimby, R.~M., Kulkarni, S.~R., Kasliwal, M.~M., et al.\ 2011, \nat, 474, 487 
\bibitem[Quimby et al.(2013)]{2013MNRAS.431..912Q} Quimby, R.~M., Yuan, F., Akerlof, C., \& Wheeler, J.~C.\ 2013, \mnras, 431, 912 
\bibitem[Quimby et al.(2018)]{2018ApJ...855....2Q} Quimby, R.~M., De Cia, A., Gal-Yam, A., et al.\ 2018, \apj, 855, 2 

\bibitem[Rakavy \& Shaviv(1967)]{1967ApJ...148..803R} Rakavy, G., \& Shaviv, G.\ 1967, \apj, 148, 803 
\bibitem[Richardson et al.(2002)]{2002AJ....123..745R} Richardson, D., Branch, D., Casebeer, D., et al.\ 2002, \aj, 123, 745 

\bibitem[Schulze et al.(2018)]{2018MNRAS.473.1258S} Schulze, S., Kr{\"u}hler, T., Leloudas, G., et al.\ 2018, \mnras, 473, 1258 
\bibitem[Smartt(2015)]{2015PASA...32...16S} Smartt, S.~J.\ 2015, \pasa, 32, e016 
\bibitem[Smith et al.(2007)]{2007ApJ...666.1116S} Smith, N., Li, W., Foley, R.~J., et al.\ 2007, \apj, 666, 1116 
\bibitem[Smith et al.(2008)]{2008ApJ...686..467S} Smith, N., Chornock, R., Li, W., et al.\ 2008, \apj, 686, 467 
\bibitem[Smith et al.(2010)]{2010ApJ...709..856S} Smith, N., Chornock, R., Silverman, J.~M., Filippenko, A.~V., \& Foley, R.~J.\ 2010, \apj, 709, 856 
\bibitem[Smith(2014)]{2014ARA&A..52..487S} Smith, N.\ 2014, \araa, 52, 487 
\bibitem[Smith et al.(2016)]{2016ApJ...818L...8S} Smith, M., Sullivan, M., D'Andrea, C.~B., et al.\ 2016, \apjl, 818, L8 
\bibitem[Smith et al.(2018)]{2018ApJ...854...37S} Smith, M., Sullivan, M., Nichol, R.~C., et al.\ 2018, \apj, 854, 37 
\bibitem[Sukhbold \& Woosley(2016)]{2016ApJ...820L..38S} Sukhbold, T., \& Woosley, S.~E.\ 2016, \apjl, 820, L38 


\bibitem[Th{\"o}ne et al.(2015)]{2015MNRAS.451L..65T} Th{\"o}ne, C.~C., de Ugarte Postigo, A., Garc{\'{\i}}a-Benito, R., et al.\ 2015, \mnras, 451, L65 

\bibitem[Valenti et al.(2008)]{2008ApJ...673L.155V} Valenti, S., Elias-Rosa, N., Taubenberger, S., et al.\ 2008, \apjl, 673, L155 
\bibitem[Vreeswijk et al.(2014)]{2014ApJ...797...24V} Vreeswijk, P.~M., Savaglio, S., Gal-Yam, A., et al.\ 2014, \apj, 797, 24 
\bibitem[Vreeswijk et al.(2017)]{2017ApJ...835...58V} Vreeswijk, P.~M., Leloudas, G., Gal-Yam, A., et al.\ 2017, \apj, 835, 58 


\bibitem[Waldman(2008)]{2008ApJ...685.1103W} Waldman, R.\ 2008, \apj, 685, 1103 
\bibitem[Wang et al.(2017)]{2017arXiv171007005W} Wang, L., Baade, D., Baron, E., et al.\ 2017, arXiv:1710.07005 
\bibitem[Woosley et al.(2007)]{2007Natur.450..390W} Woosley, S.~E., Blinnikov, S., \& Heger, A.\ 2007, \nat, 450, 390 
\bibitem[Woosley(2010)]{2010ApJ...719L.204W} Woosley, S.~E.\ 2010, \apjl, 719, L204 
\bibitem[Woosley(2017)]{2017ApJ...836..244W} Woosley, S.~E.\ 2017, \apj, 836, 244 


\bibitem[Yan et al.(2015)]{2015ApJ...814..108Y} Yan, L., Quimby, R., Ofek, E., et al.\ 2015, \apj, 814, 108 
\bibitem[Yan et al.(2017)]{2017ApJ...848....6Y} Yan, L., Lunnan, R., Perley, D.~A., et al.\ 2017, \apj, 848, 6 
\bibitem[Yan et al.(2017)]{2017ApJ...840...57Y} Yan, L., Quimby, R., Gal-Yam, A., et al.\ 2017, \apj, 840, 57 
\bibitem[Yan et al.(2018)]{2018ApJ...858...91Y} Yan, L., Perley, D.~A., De Cia, A., et al.\ 2018, \apj, 858, 91 
\bibitem[Yaron \& Gal-Yam(2012)]{2012PASP..124..668Y} Yaron, O., \& Gal-Yam, A.\ 2012, \pasp, 124, 668 

\bibitem[Yaron et al.(2017)]{2017NatPh..13..510Y} Yaron, O., Perley, D.~A., Gal-Yam, A., et al.\ 2017, Nature Physics, 13, 510 

\end{thebibliography}
\end{document}